\documentclass[twocolumn,showpacs,preprintnumbers,amsmath,amssymb]{revtex4}

\usepackage{graphicx}
\usepackage{dcolumn}
\usepackage{bm}
\usepackage{amsmath}
\usepackage{amssymb}
\usepackage{multirow}
\usepackage{textcomp}

\begin{document}

\preprint{}

\title{Two-dimensional phonon polaritons in multilayers of hexagonal boron nitride from a macroscopic phonon model}

\author{J.-Z. Zhang}
 \email{phyjzzhang@jlu.edu.cn}
 \affiliation{School of Physics, and State Key Laboratory of Superhard Materials, Jilin University, Changchun 130012, China.
}

\date{\today}
\begin{abstract}
Phonon polaritons (PhPs) in freestanding and supported multilayers (MuLs) of hexagonal boron nitride (hBN) are systematically studied using a macroscopic optical-phonon model. 
The PhP properties such as confinement, group velocity, propagation quality factor (PQF) and wavelength scaling are studied. 
Owing to the nonlocal high-frequency screening, there is an upper frequency limit  making the two-dimensional (2D) PhPs have a frequency band, and also a maximum PQF occurs near  the centre frequency.  The substrate's dielectric response should be included to accurately calculate the PhP properties. 
While the simple electrostatic approximation (ESA) is a proper treatment for PhP frequencies $\omega$ above $\omega_0$ (e.g. $\omega>1.03\omega_0$ for the 30-layers; $\omega_0$ is the $\Gamma$ point optical phonon frequency), it fails to describe the PhP properties near $\omega_0$ and the effect of retardation should be included for an accurate description.    
The PhP wavelength versus the layer thickness near $\omega_0$ deviates significantly from a linear scaling law given by the ESA due to strong phonon-photon coupling. The calculated PhP dispersion, PQF and scaling are compared with experimental data of a number of spectroscopic studies and good agreement is obtained. While  
the frequency of incident light should be near the centre frequency to maximize the PQF,   
 the PhP wavelength, confinement and propagation length can be engineered by varying the MuL thickness and its dielectric environment. 
\end{abstract}
\pacs{63.22.Np, 71.36.+c, 78.67.Pt}
\maketitle


\section{Introduction}



Phonon polaritons (PhPs) are collective modes of photons and phonons resulting from coupling light with optical lattice vibrations in a polar crystal. 
The subwavelength confinement of PhPs refers to (i) the spatial confinement to the interface quantified by the field decay length in the perpendicular direction \cite{Maier:2007} and (ii) the shortening in wavelength given by the confinement factor $\beta$,  
the ratio of the incident light wavelength to the PhP wavelength \cite{Basov:2016}. 
Hexagonal boron nitride (hBN) is a natural hyperbolic material \cite{Poddubny:2013}; 
namely, its permittivity tensor components have opposite signs.  
PhPs in bulk hBN exhibit strong field confinement at sub-diffractional length scales  \cite{Caldwell:2014,Basov:2016,Low:2017} and long lifetimes (picoseconds) \cite{Giles:2018} experimentally.  
Experiments \cite{Dai:2014,Dai:2019,Ningli:2020,Shiz:2015} have also found that the PhP confinement is enhanced in thin multilayers (MuLs) of hBN, which emerges as a promising two-dimensional (2D) material for nanoscale control of light at infrared frequencies. 

%

PhPs occur in bulk hBN with frequency $\omega$ ranging between the transverse (TO) and  longitudinal (LO) optical phonon frequencies, i.e. in the two Reststrahlen bands \cite{Dai:2014,Caldwell:2014}, where the permittivity tensor has negative in-plane ($x-y$) or out-of-plane ($z$) component(s). 
In a 2D polar monolayer (ML) such as ML hBN there are two major changes to its properties:  (i) while the TO phonons are dispersionless at long wavelengths (with frequency $\omega_0$), the LO phonons have dispersion, which are degenerate with the TO modes at the Brillouin zone centre \cite{Sanchez:2002,Michel:2009,Sohier:2017,Zhang:2020a}; (ii)
the dielectric function (DF) depends linearly on the norm of in-plane wavevector $\mathbf{k}$ (leading to nonlocal screening) \cite{Cudazzo:2011}, e.g., the high-frequency DF $\epsilon_{\infty}(k)=1+2\pi\chi_ek$ for an isolated ML, where $\chi_e$ the ML's electronic susceptibility (i.e. high-frequency susceptibility). Using the conductivity as optical response function,  
 a latest theoretical study \cite{Rivera:2019} has found that despite no LO-TO splitting at $\Gamma$, highly confined PhPs occur in a polar ML as transverse magnetic waves, which are treated there as the 2D LO phonons using the electrostatic approximation (ESA)  \cite{Rivera:2019}.   
In conductivity $\sigma(\mathbf{k},\omega)$ the  latter term of $\epsilon_{\infty}(k)$, i.e. the high-frequency screening (HFS) due to  $\chi_e$   
 is neglected \cite{Rivera:2019}.  Surely the nonlocal HFS is very weak for small-wavevector PhPs in an ML; 
for deep subwavelength PhPs of interest, 
 $k\gg1/2\pi\chi_e$, however the HFS can significantly affect the PhP properties and its influence becomes stronger in an MuL due to the increased dielectric susceptibility. 
 PhPs at a large $k$, $k\gg\omega_0/c$ ($c$ is the speed of light in vacuum), can be evaluated usually with a simpler ESA \cite{Rivera:2019} as the retardation effect (RE) (i.e. ion-ion interactions propagate with the finite speed $c$) is very small, but the small-$k$ PhPs (with $\omega\sim\omega_0$) need to be calculated rigorously from   Maxwell's equations to include the strong RE \cite{Born:1954}. 
So far, there is a lack of theoretical studies of 2D PhPs and the effects of HFS and retardation are unclear.   
 
Experimentally, PhPs in bulk (three dimensions, 3D) \cite{Giles:2018}, bulk and MuLs \cite{Dai:2014,Shiz:2015}, MLs and bilayers \cite{Dai:2019} of hBN on a SiO$_2$ substrate have been studied using scattering scanning near-field optical microscopy (s-SNOM). In a latest study PhP dispersions have been measured also for {\it freestanding} hBN MLs and MuLs using electron energy-loss spectroscopy (EELS) \cite{Ningli:2020}, with very strong PhP confinement ($\beta>487$) and very small group velocity $\sim$$10^{-5}c$ obtained in the MLs. 
Both experiments \cite{Dai:2014,Shiz:2015} have found that for a high incident light  frequency the PhP wavelength increases proportionally with the layer thickness while the wavelength scaling deviates from the linear relation at lower frequencies. Theory predicts recently  that the product of 2D wavevector $k$ and the number of layers $\mathcal{N}$ depends on the LO mode frequency \cite{Sohier:2017}, suggesting that the linear scaling in MuLs may be due to the LO phonon character in the PhPs, which however needs to be examined quantitatively.  
Further the observed nonlinear scaling has yet to be resolved.

Recently a macroscopic model 
including electronic polarization of ions (i.e. HFS) and local field effects has been used to study  
optical phonons in ML hBN \cite{Zhang:2020a}. In this paper, we extend this model to $\mathcal{N}$ layers of hBN to study PhPs due to the in-phase optical vibrations  (i.e. the ionic motions in different layers are in-phase) \cite{Sohier:2017,Michel:2011}. 
The purpose of this study is twofold: (i) to derive PhP modes including both HFS and REs and then evaluate the influence of the nonlocal HFS and examine the nonretarded approximation (i.e. the ESA) for the PhP calculations, and (ii) to study PhP properties such as the confinement, group velocity, PQF and wavelength scaling, and also use them to analyze the experimental results and find out the causes of the observed linear and nonlinear scaling.  
The dielectric environment (DE) effects are obtained by comparing the PhP dispersions in freestanding and supported MuLs and also by comparing results calculated with various dielectric constants (DCs) of the substrate.   
When accounting for the HFS and REs we obtain qualitatively different results of the key PhP properties and also good agreement with the experiments.     

The content of this paper is as follows. In Section II, a macroscopic model for in-phase optical vibrations in $\mathcal{N}$-layer hBN is described and then LO mode and PhP dispersion relations are derived including the HFS and DE effects.   
In Section III, we begin with the optical response and absorption spectra to show  the PhP frequencies being limited due to the HFS. Next 
we show numerical results of PhP dispersion ($\beta$ versus $\omega$) in freestanding and on-SiO$_2$ hBN MuLs,     
to evaluate the effects due to the HFS, retardation and DE,  
and then compare the calculated dispersions with experiment.  
We also present the PhP properties, i.e. the group velocity, density of states (DOS) and decay and propagation lengths versus frequency and their variation with $\mathcal{N}$ and the DE.  We compare the calculated PhP PQFs of $\mathcal{N}$-layer and bulk hBN, and also compare with the PQF data found from bulk measurements.   
Further we present wavelength scaling at various incident frequencies and compare with the experiments. 
 Finally, Section IV summarizes the main results obtained. 
In Appendix A, transparent expressions are derived for the PhP dispersion and scaling near the TO phonon frequency, which are used to interpret the numerical results due to the  retardation. In Appendix B,  
 a PhP group velocity expression including the full dielectric response is obtained and used for accurate calculations while       
simplified ESA expressions are given to analyze the numerical group velocities and DOS. 
 

\section{Theory}

\subsection{Macroscopic model for in-phase optical vibrations in an MuL}

In a latest study \cite{Zhang:2020a}, the long wavelength in-plane optical vibrations of ML hBN are described by a pair of macroscopic equations,
$\ddot{\mathbf{w}}=a_{11}\mathbf{w}+a_{12}\mathbf{E}_{\boldsymbol{\rho}}$, and   
$\boldsymbol{\mathcal{P}}=a_{21}\mathbf{w}+a_{22}\mathbf{E}_{\boldsymbol{\rho}}$, where   $\mathbf{w}$ is the optical displacement between the positive and negative ions weighted by $\sqrt{\bar{m}/s}$, $\bar{m}$ being their reduced mass and $s$ the unit-cell area,  $\boldsymbol{\mathcal{P}}$ is the macroscopic polarization (average dipole moment per unit {\it area}), and  $\mathbf{E}_{\boldsymbol{\rho}}$ the in-plane ($\boldsymbol{\rho}$) component of the {\it macroscopic} field in the ML. All in-plane vectors $\mathbf{w}$, $\boldsymbol{\mathcal{P}}$ and $\mathbf{E}_{\boldsymbol{\rho}}$ depend on time $t$ and in-plane position vector $\boldsymbol{\rho}$. The local field effects \cite{Born:1954} are included via the $a$-coefficients so that the field appearing in the macroscopic equations  is only the macroscopic field.  
The three independent $a$-coefficients correspond to three independent first-principles calculated quantities, namely, (i) $\omega_0$, the intrinsic oscillator frequency \cite{Born:1954}, equal to the TO phonon frequency, (ii) $e_B$, the Born charge \cite{Gonze:1997} of positive ions B, and (iii) $\chi_e$, the {\it electronic} susceptibility of the 2D material, via  
$a_{11}=-\omega_0^2$, 
$a_{12}=a_{21}=e_B/\sqrt{\bar{m}s}$, and 
$a_{22}=\chi_e$. CGS units are used throughout the paper. Evidently 
the in-plane lattice susceptibility of the ML  $\chi=\boldsymbol{\mathcal{P}}/\mathbf{E}_{\boldsymbol{\rho}}$ at perturbing frequency $\omega$ is   
\begin{equation}
	\chi(\omega)=\chi_e+\frac{e_B^2/(\bar{m}s)}{\omega_0^2-\omega^2},               
	\label{chiml}
\end{equation}
where $\chi_e$ is the high-frequency susceptibility ($r_{eff}=2\pi\chi_e$ is an effective screening length \cite{Cudazzo:2011}), and the $\omega$-dependent term is due to lattice vibrations,  
$e_B^2/\bar{m}$ being the mode-oscillator strength \cite{Gonze:1997}. 

Bulk hBN is a layered van der Waals crystal. For $\mathcal{N}$ layers of hBN, modeled as a dipole lattice \cite{Born:1954,Zhang:2019b}, 
our calculation finds that the field at an ion site due to the dipoles in an adjacent layer is two orders of magnitude smaller than the field of dipoles in the same layer. Thus for the dipole field acting on each ion we consider only the dipoles of the same layer as we did for ML hBN \cite{Zhang:2019b}. Then following the derivation in Ref. \cite{Zhang:2019b}, we readily find that the macroscopic equations above can describe the lattice motions of each layer in the MuL, with all quantities now dependent on the layer; for instance, the in-plane field is $\mathbf{E}_{\boldsymbol{\rho}}(\boldsymbol{\rho},z_i)$, $z_i$ being the layer's $z$-coordinate.    
In consequence,  
the $\mathcal{N}$-layer is treated, physically speaking, as $\mathcal{N}$ single layers which are however coupled due to the {\it macroscopic} field.  
The macroscopic field is considered uniform over the MuL, i.e. $\mathbf{E}_{\boldsymbol{\rho}}(\boldsymbol{\rho},z_i)$ is independent of $z_i$  \cite{Sohier:2017}. In a perturbing in-plane field of frequency $\omega$, 
all layers vibrate and the ionic displacements (thus the polarization) in different layers are the same, i.e. in-phase \cite{Michel:2011,Sohier:2017}, so that the polarization from each layer adds up to give the total macroscopic polarization $\boldsymbol{\mathcal{P}}_{\mathcal{N}}$. 
Thus $\mathcal{N}$ layers of hBN has 
an in-plane lattice susceptibility $\chi_{\mathcal{N}}=\boldsymbol{\mathcal{P}}_{\mathcal{N}}/\mathbf{E}_{\boldsymbol{\rho}}$ given by $\mathcal{N}$ times the ML susceptibility $\chi$ [Eq.~(\ref{chiml})],  
$\chi_{\mathcal{N}}(\omega)=\mathcal{N}\chi(\omega)$; that is, both the screening length and the mode-oscillator strength are multiplied by $\mathcal{N}$ \cite{Sohier:2017}. 
  In deriving the LO and PhP modes below, the MuL is treated as a 2D dielectric with a surface {\it polarization charge} density entering the boundary condition on electric displacement $\mathbf{D}$.

\subsection{In-phase LO phonons in an MuL}

The in-phase LO modes are connected with a  macroscopic field and 
 correspond to the highest LO branch of an hBN MuL  \cite{Michel:2011,Sohier:2017,Dai:2014,Dai:2019};  $\mathbf{w}\parallel\mathbf{E}_{\boldsymbol{\rho}}\parallel\boldsymbol{\mathcal{P}}\parallel\mathbf{k}$ in all layers,  $\mathbf{k}$ being the 2D wavevector. 
We consider a general case, an $\mathcal{N}$-layer hBN embedded between two bulk crystals with lattice DFs $\epsilon_1(\omega)$ ($z<0$; e.g., the substrate) and $\epsilon_2(\omega)$ ($z>0$).

The equation of electrostatics is   
$\nabla\cdot\mathbf{D}=\nabla\cdot[\epsilon(\omega)\mathbf{E}]=0$, where the electrostatic field $\mathbf{E}=-\nabla\phi$, and $\epsilon(\omega)=\epsilon_1(\omega)$ [$\epsilon_2(\omega)$] in the lower (upper) half space $z<0$ ($z>0$). Let $\mathbf{k}$ along the $x$-axis and the electric potential is  $\phi(x,z)=\varphi(z)e^{ikx-i\omega t}$  
due to in-plane isotropy (similarly write $\mathbf{E}$ and $\mathbf{D}$). 
Consider frequencies such that $\epsilon_1(\omega)\ne 0$ and $\epsilon_2(\omega)\ne 0$. The equation of electrostatics simplifies to Laplace's equation $\nabla^2\phi=0$, with solutions  $\varphi(z)=Ae^{kz}$ and $\varphi(z)=Be^{-kz}$ in the lower and upper media, respectively.  
As the in-plane electric field is continuous across the interface, one has $A=B$ and also the surface polarization charge per unit area  $\sigma_p=-\nabla_{\boldsymbol{\rho}}\cdot \boldsymbol{\mathcal{P}}_{\mathcal{N}} =-\nabla_{\boldsymbol{\rho}}\cdot[\mathcal{N}\chi(\omega)\mathbf{E}_x(x,0)]$.
Using the boundary condition on $\mathbf{D}$, 
$D_z(x,z\rightarrow0^+)-D_z(x,z\rightarrow0^-)=4\pi\sigma_p$, then one finds    
\begin{equation}
\epsilon_1(\omega)+\epsilon_2(\omega)=-4\pi\mathcal{N}k\chi(\omega),               
\label{lonlay0}
\end{equation}
the solution of which gives the LO phonon dispersion $\omega(k)$ of $\mathcal{N}$-layer hBN in the two crystals. In an LO mode, evidently there are electrostatic potentials and fields in both crystals as well as the hBN MuL. 
Without hBN layers Eq.~(\ref{lonlay0}) reduces to $\epsilon_1(\omega)+\epsilon_2(\omega)=0$, 
yielding frequencies of the interface phonons of the heterostructure constituted of the two crystals \cite{Zhang:2011}.  
For TO vibrations in the $\mathcal{N}$-layer, $\mathbf{w}\perp\mathbf{k}$, $\boldsymbol{\mathcal{P}}\perp\mathbf{k}$ in all layers; there is no surface polarization charge,  $\sigma_p=-\nabla_{\boldsymbol{\rho}}\cdot \boldsymbol{\mathcal{P}}_{\mathcal{N}}=0$. Then one finds that $A=B=0$, and the macroscopic field vanishes,  $\mathbf{E}(\mathbf{r})=0$. Thus the TO modes are dispersionless with frequency $ \omega_t=\omega_0$, equal to that of ML hBN.  
When $\epsilon_1(\omega)=0$ ($\epsilon_2(\omega)=0$), the vibrations correspond to half-space LO phonons of the lower (upper) crystal \cite{Zhang:2011} with electric potentials and fields occurring only in that half-space, which are not considered  here. 
DFs $\epsilon_1(\omega)$ and $\epsilon_2(\omega)$ are usually treated as constants, giving  background dielectric constant (DC) $\epsilon_b=(\epsilon_1+\epsilon_2)/2$ \cite{Sohier:2017}. 
 
For a freestanding $\mathcal{N}$-layer [$\epsilon_1(\omega)=\epsilon_2(\omega)=1$], inserting expression (\ref{chiml}) into Eq.~(\ref{lonlay0}) yields the LO phonon frequency 
\begin{equation}
\omega^2=\omega_0^2+\frac{2\pi\mathcal{N}e_B^2k/(\bar{m}s)}{1+2\pi\mathcal{N}\chi_ek},     \label{wlo2frea1}
\end{equation}
which is identical to the expression  $\omega_l^2=\omega_0^2+\mathcal{N}\mathcal{S}k/(1+\mathcal{N}r_{eff}k)$ given in a recent study  [i.e., Eq.~(5) of Ref.\cite{Sohier:2017}], where $\mathcal{S}$ is the LO-TO splitting strength in the ML $\mathcal{S}=2\pi e_B^2/(\bar{m}s)$.  The denominator is the high-frequency DC corresponding to the nonlocal HFS. 
This LO mode dispersion is in good agreement with calculations by density-functional perturbation theory and  describes the layer number ($\mathcal{N}$) dependence of the LO-TO splitting  \cite{Michel:2011,Sohier:2017}. The zone center LO mode group velocity increases proportionally with $\mathcal{N}$ [consistent with the previous result, Eq.~(59b) of Ref.\cite{Michel:2011}], $v_{l,\mathcal{N}}=\mathcal{N}v_l$, where $v_l$ is the group velocity in an ML, $v_l=\pi e_B^2/(\bar{m}s\omega_0)$.   
These LO phonon results will be used below (section III) to examine the ESA and analyze retardation effects in PhP modes.


\subsection{In-phase phonon polaritons in an MuL}


We confine ourselves to transverse magnetic PhPs \cite{Maier:2007,Rivera:2019}; i.e.,   
 the magnetic field $\mathbf{H}$ is parallel to the $\mathcal{N}$ {\it layers} of hBN and perpendicular to $\mathbf{k}$ while the electric field's in-plane component is along $\mathbf{k}$, leading to  $\mathbf{w}\parallel\mathbf{E}_{\boldsymbol{\rho}}\parallel\boldsymbol{\mathcal{P}}\parallel\mathbf{k}$ in {\it each layer} as in the LO modes. 
These in-phase PhPs are connected with the {\it macroscopic} polarization charge and current and thus correspond to the strongly confined electromagnetic modes measured in experiments (see Supplementary Materials of Ref.\cite{Dai:2014}).  
 Let the PhP waves propagate in the $x$ direction, $\mathbf{E}=\mathbf{E}(z)e^{ikx-i\omega t}$ and 
$\mathbf{H}=H(z)e^{ikx-i\omega t}\mathbf{e}_y$,  where $E_x$, $E_z$ and $H$ are nonzero  ($\mathbf{H}\parallel\mathbf{e}_y$).  
Solving the wave equation of $\mathbf{H}$ i.e.  $\nabla^2\mathbf{H}=\epsilon(\omega)\ddot{\mathbf{H}}/c^2$ yields  
$H(z)=Fe^{K_1z}$ and $H(z)=Ge^{-K_2z}$ in the lower and upper crystals, where 
$K_1=\sqrt{k^2-\epsilon_1(\omega)\omega^2/c^2}$, and  
$K_2=\sqrt{k^2-\epsilon_2(\omega)\omega^2/c^2}$. 
From the equation $\nabla\times\mathbf{H}=\epsilon(\omega)\dot{\mathbf{E}}/c$ one finds
$E_x(z)=-icH'(z)/(\omega\epsilon(\omega))$ and  
$E_z(z)=-ckH(z)/(\omega\epsilon(\omega))$. 
Continuity of $\mathbf{E}_x$, $E_x(x,z\rightarrow 0^-)=E_x(x,z\rightarrow 0^+)$, leads to $K_1F/\epsilon_1(\omega)=-K_2G/\epsilon_2(\omega)$, and  
the boundary condition on $\mathbf{D}$  
 yields the expression $F-G=4\pi\mathcal{N}\chi(\omega)K_2G/\epsilon_2(\omega)$. Therefore the PhP dispersion is given by 
\begin{equation}
\frac{\epsilon_1(\omega)}{K_1}+\frac{\epsilon_2(\omega)}{K_2}
=-4\pi\mathcal{N}\chi(\omega).  
\label{polaton0}
\end{equation}

A similar equation containing conductivity $\sigma$ has been obtained for ML hBN recently  \cite{Rivera:2019} [$\sigma$  relates to the ML susceptibility via $\sigma(\omega)=-i\omega\chi(\omega)$]. Apart from 2D ionic vibrations, high-frequency susceptibility $\chi_e$ also contributes to $\chi(\omega)$ and $\sigma(\omega)$ [see Eq.~(\ref{chiml})]; however the HFS was neglected in Ref.\cite{Rivera:2019}. 
At large $k$ PhP phase velocity $v\ll c$ the ion-ion interaction can be treated with the unretarded Coulomb force, i.e., the ESA \cite{Born:1954}. 
Then one has $K_1\approx K_2\approx k$ and the dispersion equation (\ref{polaton0}) reduces to a simpler LO mode dispersion formula [Eq.~(\ref{lonlay0})] \cite{Rivera:2019}. 
 Eq.~(\ref{polaton0}) simplifies to the equation for surface plasmon polaritons \cite{Maier:2007} at a metal-dielectric interface when letting $\chi(\omega)=0$ and using the Drude DF \cite{Haug:2004} for the metal $\epsilon_1(\omega)$. 


 


The confinement factor $\beta$ is a key polariton property   \cite{Basov:2016,Low:2017,Pli:2016,Dai:2019,Rivera:2019,Ningli:2020}, $\beta=k/k_a$, where $k_a$ is the corresponding photon wavevector in vacuum, $k_a=\omega/c$.  Evidently $\beta=\lambda_a/\lambda=c/v$, where $\lambda$ and $\lambda_a$ are the polariton wavelength and its corresponding vacuum wavelength. The polariton dispersion relation is usually given by the dependence of the confinement factor $\beta$  on frequency $\omega$ \cite{Pli:2016,Low:2017,Dubrovkin:2018,Dai:2019,Ningli:2020}. Photons in a medium with DC $\epsilon$ have dispersion  $\omega=ck/\sqrt{\epsilon}$, which is represented by $\beta=\sqrt{\epsilon}$ in the ($\omega$,$\beta$) plot.  Let $k_0$ be the vacuum wavevector corresponding to phonon frequency $\omega_0$, $k_0=\omega_0/c$. Two dimensionless quantities  
$\alpha=k_0\chi_e$ and $\eta=v_l/(\pi\omega_0\chi_e)$ are frequently used below. Using $\omega_0$ and $k_0$ as the units of frequency $\omega$ and wavevector $k$, $y=\omega/\omega_0$, $x=k/k_0$, 
Eq.~(\ref{polaton0}) can be nondimensionalized (Appendix A), and the confinement factor is  simply $\beta=x/y$.  Given $\omega$, $\beta$ is obtained analytically for freestanding layers from Eq.~(\ref{pofre5ap}).  For hBN layers on a substrate generally with DF $\epsilon_1(\omega)=\bar{\epsilon}_1(y^2)$ (see Appendix A), the PhP dispersion and scaling need to be calculated numerically by solving Eq.~(\ref{pofre2ap}) with a standard root-finding method.

 To evaluate polariton propagation we need to include phonon damping. A force opposing the motion  $-\gamma\dot{\mathbf{w}}$ \cite{Born:1954} can be added to the former macroscopic equation, $\gamma$ being the damping rate, and accordingly an additional term $-i\gamma\omega$ will enter  the denominator of the susceptibility Eq.~(\ref{chiml}). Now the susceptibility $\chi(\omega)$ in the ESA Eq.~(\ref{lonlay0}) and PhP  dispersion Eq.~(\ref{polaton0}) becomes a complex function, and the PhP wavevector $k$ becomes a complex number. 
As the light intensity $I\propto\lvert\mathbf{E}\rvert^2$ we introduce the lattice absorption coefficient for $\mathcal{N}$-layer hBN 
as $\alpha(\omega)$=2Im($k$) as for bulk crystals \cite{Born:1954,Haug:2004}; the reciprocal of $\alpha(\omega)$ gives the {\it propagation length} over which $I$ decreases by a factor $e^{-1}$.  
The propagation quality factor (PQF) is given by $\gamma_p^{-1}$=$\lvert$ Re($k$)/Im($k$)$\rvert$, which represents the figure 
of merit for polariton propagation losses \cite{Basov:2016,Low:2017,Rivera:2019,Dubrovkin:2018,Giles:2018,Dai:2014,Dai:2019}.

The lattice constant of ML hBN is $a$=2.5 $\AA$, giving a unit-cell 
area $s=\sqrt{3}a^2/2$=5.4127 $\AA^2$. 
The masses of the boron and nitrogen atoms are $m_1$=10.811 Da and $m_2$=14.0067 Da, yielding a reduced mass $\bar{m}$=6.1015 Da.  From the first-principles calculated ML quantities $\mathcal{S}=8.4\times 10^{-2}$ eV$^2\cdot$$\AA$ and $r_{eff}$=7.64 $\AA$ \cite{Sohier:2017} one finds that  $e_B=2.71e$ and $\chi_e$=1.216 $\AA$.  
The frequency $\omega_0$=169.98 meV calculated in Ref.\cite{Erba:2013}, very close to other first-principles \cite{Wirtz:2003,Serrano:2007} and also experimental  \cite{Rokuta:1997,Dai:2014,Caldwell:2014,Dai:2019,Shiz:2015} values, corresponds to a wavevector $k_0=8.62\times10^{-5}/\AA$.     
 Thus the LO phonon group velocity $v_l$=37.54 km/s, and the  
two dimensionless quantities  
$\alpha\approx0.0001$ and $\eta\approx 0.38$. In the PhP dispersion calculations the wavevectors are restricted to a small-$k$ region, $k<0.1\lvert \Gamma-K\rvert$ ($\lvert \Gamma-K\rvert$ is the distance between points $\Gamma$ and $K$ in the Brillouin zone), where the macroscopic description of the polar modes is found to be accurate \cite{Sohier:2017,Zhang:2020a}.  These parameters are used throughout this paper.

\section{Numerical results and discussions}

\subsection{PhPs in freestanding hBN $\mathcal{N}$-layers}

\begin{figure}
	\includegraphics*[width=8cm]{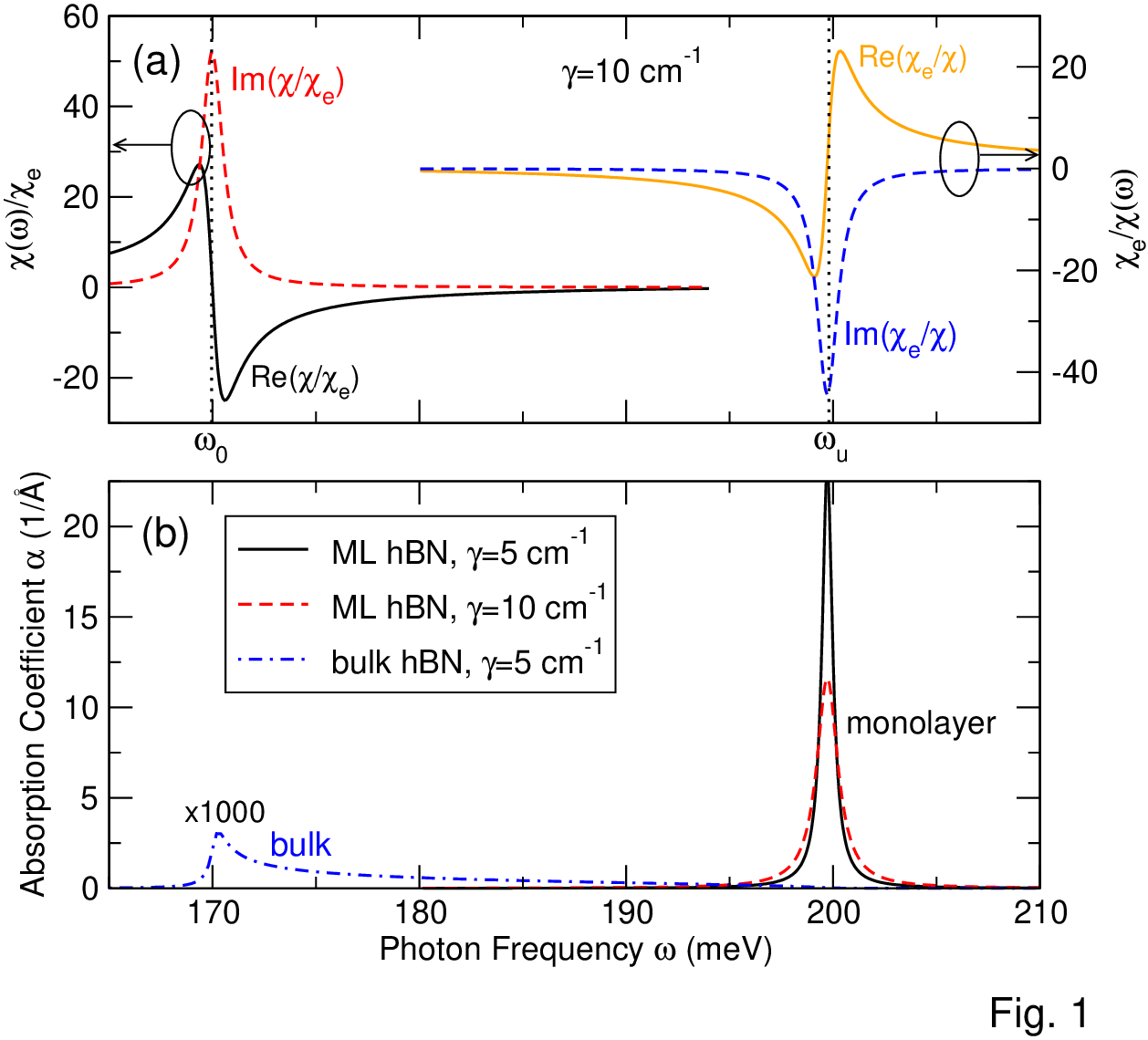}
	\caption
	{(Color online) (a) Real and imaginary parts of the susceptibility $\chi(\omega)$ (left vertical axis) and its inverse $1/\chi(\omega)$ (right vertical axis) with damping rate $\gamma$=10 cm$^{-1}$ of ML hBN. The PhP band is from $\omega_0$ (left vertical dotted line) to $\omega_u$ (right vertical dotted line). (b) Lattice absorption coefficients of ML hBN versus photon frequency $\omega$ for two damping rates $\gamma=$5 and 10 cm$^{-1}$ (the absorption peaks at $\omega_u$) and the bulk hBN absorption spectrum  (dot-dashed line, enlarged by 1000 times; $\gamma=$5 cm$^{-1}$).  
	}
	\label{fig1}

\vspace*{2cm}



	\includegraphics*[width=8cm]{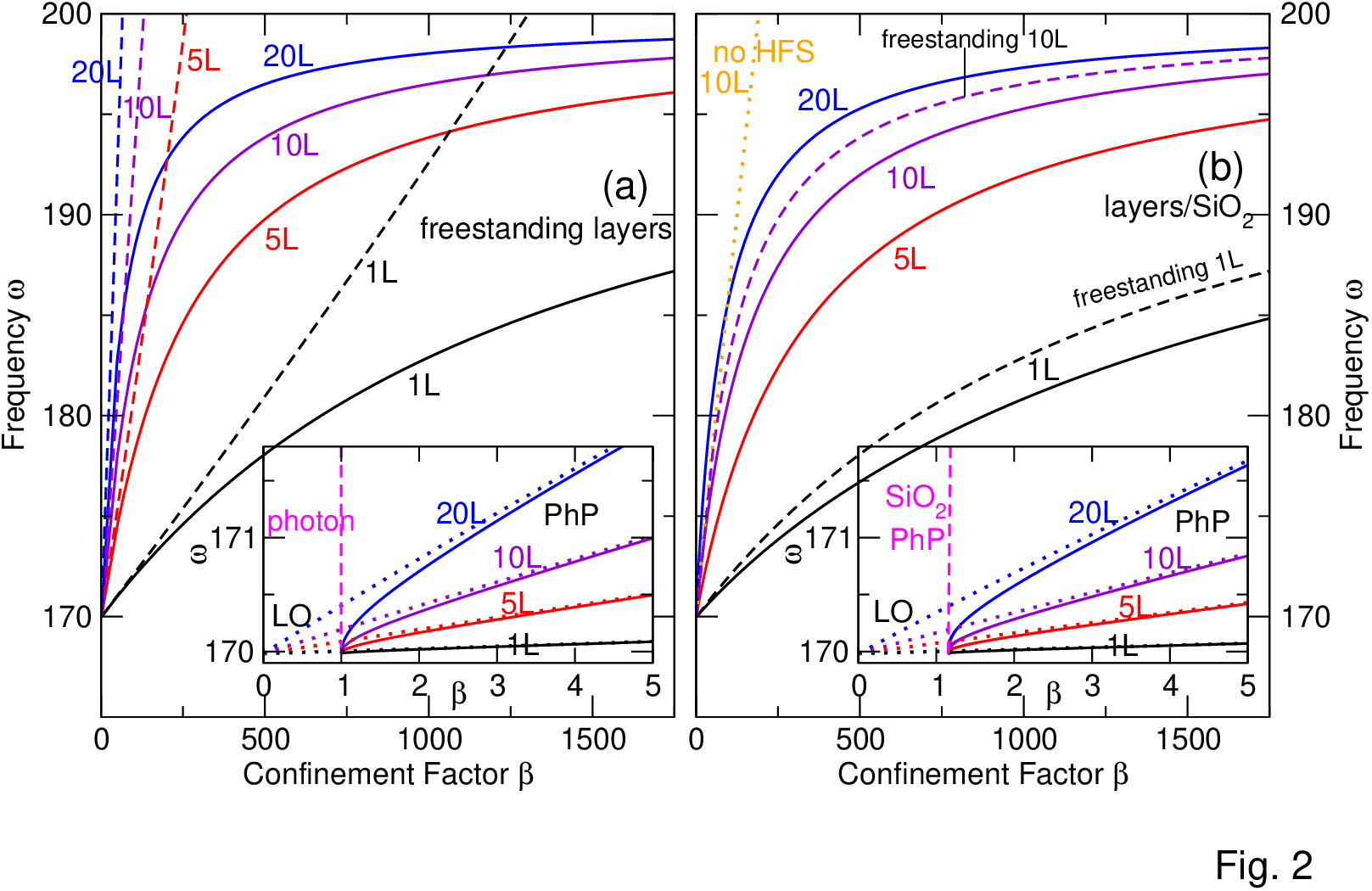}
	\caption
	{(Color online) Dispersion relations of PhPs (confinement factor versus frequency) in (a) four different freestanding hBN layers, i.e., an ML (1L), 5 layers (5L), 10 layers (10L) and 20 layers (20L) of hBN, and (b) the four hBN layers on a SiO$_2$ substrate calculated accounting for the HFS (solid lines). The results without HFS are also shown in (a) for all layers (dashed lines) and (b) for the 10L (dotted line); the freestanding 1L and 10L results including HFS are also added to (b) (dashed lines). The inset in (a) [(b)] shows the strong RE and phonon-photon coupling by comparing the dispersion relations of 2D PhPs (solid curves), LO phonons (dotted lines) and free-space photons (vertical dashed line)  [PhPs of bulk SiO$_2$ (dashed line, not exactly vertical)] near the minimum frequency $\omega_0$. 
	}
	\label{fig2}
\end{figure}

The ML susceptibility $\chi(\omega)$ [Eq.~(\ref{chiml})] is a key response function  determining the PhP dispersion [Eq.~(\ref{polaton0})]. 
Figure \ref{fig1}(a) shows $\chi(\omega)$ and its inverse $1/\chi(\omega)$ with a damping rate $\gamma$=10 cm$^{-1}$ for ML hBN. 
Evidently $\chi(\omega)<0$ is required for the PhPs, corresponding physically to 2D polarization $\boldsymbol{\mathcal{P}}$ being  antiparallel to electric field  $\mathbf{E}_{\boldsymbol{\rho}}$. Therefore,  
 the PhP frequencies of an $\mathcal{N}$-layer are limited in the {\it range} $\omega_0<\omega<\omega_u$, where $\omega_u=\omega_0\sqrt{1+\eta} 
 \approx1.175\omega_0\approx199.722$ meV; $\omega_0$ and $\omega_u$ are both in the mid-infrared region, corresponding to the vacuum wavelengths 7.2 nm and 6.2 nm,    respectively.  The centre frequency is $\omega_c\approx1.087\omega_0\approx184.852$ meV.  This frequency band is marked in Fig.~\ref{fig1}(a), where  
$\omega_0$ is the pole of $\chi(\omega)$ (left vertical dotted line) whereas $\omega_u$ is the pole of $1/\chi(\omega)$ (right vertical dotted line). The nonlocal HFS (i.e. $\chi_e>$0) is essential to a finite {\it upper} bound $\omega_u$ for PhP frequency; neglecting it causes $\omega_u\rightarrow\infty$ (see Ref.\cite{Rivera:2019} or our result Fig.~\ref{fig2} below). 
The {\it inverse} susceptibility $1/\chi(\omega)$ becomes physically more significant since with the ESA Eq.~(\ref{lonlay0}) its imaginary part determines the lattice absorption coefficient of ML hBN, $\alpha(\omega)$=-Im$(1/\chi(\omega))/(\mathcal{N}\pi)$ [see Fig.~\ref{fig1}(b)]. The $\mathcal{N}$-layer absorption coefficient decreases simply by a factor of $1/\mathcal{N}$. 
 Different from the 2D case, 
the bulk absorption coefficient $\alpha_{3D}(\omega)$ is proportional to Im$(\chi_{3D}(\omega))$, where the susceptibility $\chi_{3D}(\omega)$ has a pole at TO mode frequency $\omega_0$ \cite{Born:1954,Haug:2004}. While the 3D hBN absorption peaks at $\omega_0$ [Fig.~\ref{fig1}(b)], the 2D layer absorption peaks not at $\omega_0$ but at $\omega_u$, the upper frequency limit, and the ML absorption is three orders of magnitude stronger than the bulk absorption \footnote[1]{In the bulk hBN calculations of this study [Fig.~\ref{fig1}(b), Fig.~\ref{fig7} and Fig.~\ref{fig8}(c)], an LO-TO type of lattice DF is used with the parameters taken from Ref.\cite{Cai:2007}, as in Ref.\cite{Dai:2014} (see its Supplementary Materials). }.

Figure \ref{fig2}(a) shows the PhP  
dispersion relations of four freestanding hBN crystals with different numbers of layers, ML (1L), 5 layers (5L), 10 layers (10L) and 20 layers (20L), calculated with or without including HFS to study the HFS effects. 
With no HFS ($\chi_e$=0), confinement factor $\beta$ increases approximately linearly with  frequency $\omega$ (dashed lines), both having no finite upper limit.  Including nonlocal HFS causes a faster increase in $\beta$ (solid lines) and restricts frequency $\omega$ to below $\omega_u$ (199.722 meV). Given a frequency, $\beta$ is underestimated when neglecting HFS, and  
the underestimate, given by the decrease of the confinement factor divided by the $\beta$ including the HFS, increases as the frequency becomes larger. The coefficients in Eq.~(\ref{pofre5ap}) (including HFS) and Eq.~(\ref{taylor2ap}) (neglecting HFS) $1/(2\pi \mathcal{N}\alpha)^2\sim10^6$ and  $[c/(2\mathcal{N}v_l)]^2\sim10^7$ e.g. for an ML are very large when $\omega>1.001\omega_0$, 
making the last term much greater than the $y^2$ term, and therefore  
the underestimate is simply $[(\omega/\omega_0)^2-1]/\eta$, {\it independent} of layer number $\mathcal{N}$. For $\omega$ near  $\omega_0$, the influence of the HFS is very small with the underestimate of $\beta$ given by $[c/(2\mathcal{N}v_l)]^2[(\omega/\omega_0)^2-1]^3/\eta$. 
There is a 48\% underestimate at the centre frequency $\omega_c$ $\sim$185 meV, and the underestimate of $\beta$ rises to 83\% at $\omega$=195 meV and approaches 1 near upper bound $\omega_u$. In what follows the HFS is included unless otherwise stated.

 Near $\omega_0$ the dispersions of the PhPs and LO phonons (i.e. $c\rightarrow\infty$) are very different, as shown in the inset of Fig.~\ref{fig2}(a) for $\beta\le 5$ (recall that abscissa $\beta$ is the ratio of wavelengths, $1/\beta=\lambda/\lambda_a$).   This is a long-wavelength region where the wavelengths $\lambda$ are not much shorter than the typical value $2\pi c/\omega_0\sim\lambda_a$ \cite{Huang:1951} and the strong phonon-photon coupling occurs. 
 The ESA is poor since the lattice vibrations are significantly affected by the retardation.  
The LO modes [Eq.~(\ref{wlo2frea1})] have the minimum frequency $\omega_0$ at $k$=0 \cite{Michel:2011,Sohier:2017,Zhang:2020a}. For PhPs,  however the same minimum frequency occurs not at $\Gamma$ but at a {\it finite} wavevector $k_0=\omega_0/c\approx8.62\times10^{-5}/\AA$ (Appendix A);  accordingly $\beta$=1. No PhP modes exist below the minimum wavevector $k_0$, i.e., $0<k<k_0$, different from the usual bulk PhPs \cite{Born:1954} and surface plasmon polaritons \cite{Maier:2007,Abajo:2010} which have no wavevector gap. As $\omega\rightarrow\omega_0$ the PhPs of the $\mathcal{N}$-layers behave like photons  (Appendix A), having dispersion $\omega=ck$ and a phase velocity and a group velocity [Fig.~\ref{fig3}(a)] both equal to the speed of light $c$, independent of layer number $\mathcal{N}$. The dispersion relation of photons, $\omega=ck$, is represented by $\beta=1$, i.e., the vertical dotted line in the ($\omega$,$\beta$) plot [inset of Fig.~\ref{fig2}(a)]. The PhP dispersion curves lie to the right of the photon line ($\beta=1$) due to the bound nature of PhPs. As $\omega$ increases 
the PhP dispersion ($\beta$ versus $\omega$) near $\omega_0$ is approximated by Eq.~(\ref{dispbeta}).  


As $1/\beta=v/c$, $1/\beta$ is used conveniently to quantify the RE. The ESA expression (\ref{wlo2frea1}) is a good approximation for the PhP dispersion when  $\omega$ is above  lower bound $\omega_0$,  for instance, when $\omega>1.01\omega_0$ for 10-layer hBN ($\omega>1.001\omega_0$ for MLs and $\omega>1.03\omega_0$ for 30-layers), as the  corresponding $1/\beta$ is very small, $1/\beta<0.1$. The LO mode dispersion curves coincide with the solid curves if plotted in Fig.~\ref{fig2}(a).  In fact, the LO phonon group velocity $v_l$ is much smaller than $c$, $v_l/c=1.25\times 10^{-4}$, corresponding literally to a flat LO mode dispersion curve \cite{Sohier:2017,Zhang:2020a} when compared to the photon line $\omega=ck$. Owing to the enormously large coefficient $[c/(2\mathcal{N}v_l)]^2$ in Eq.~(\ref{taylor2ap}) a small increase of $y$ (i.e. $\omega/\omega_0$) from 1 causes a large increase to $x^2$ thus rapidly decreasing the PhP phase velocity $v$ 
($v=cy/x$) and consequently the RE. For instance, a 1\textperthousand~increase in $\omega/\omega_0$ reduces the $v$ in MLs to $\sim$0.1$c$. Away from $\omega_0$, as phase velocity $v$ decreases, the RE becomes much weaker and therefore the ESA result is much closer to the PhP dispersion. However 
the ESA becomes more inaccurate near $\omega_0$ in a {\it thicker} $\mathcal{N}$-layer [inset of Fig.~\ref{fig2}(a)] where the phonon-photon interaction is stronger. 
 In terms of wavevectors the ESA is applicable in the $\mathcal{N}$ layers for $k$ far above the wavevector gap, $k>10k_0\approx0.0005\lvert \Gamma-K\rvert$. 
 Having this, 
 we can estimate wavevectors [using Eq.~(\ref{wlo2frea1})] for which the HFS can be neglected; clearly they are a small portion near $\Gamma$ (a smaller portion in a thicker layer), i.e. $k\ll 1/(2\pi\mathcal{N}\chi_e)$, with $1/(2\pi\chi_e)\approx0.08\lvert \Gamma-K\rvert$. 


The group velocity $v_g$ corresponds to the slope of the dispersion and is a key PhP property \cite{Basov:2016,Low:2017,Dai:2015,Caldwell:2014,Giles:2018,Ningli:2020}. 
At $\omega_0$, $v_g$ approaches $c$ while the PhP DOS $g_p$ approaches the 2D photon DOS $\omega_0/(2\pi c^2)$ for all $\mathcal{N}$-layers (Appendix B).  
With a small increase of $\omega$, then $v_g$ falls dramatically while $g_p$ increases rapidly [Fig.~\ref{fig3}(a)] owing to the strong phonon-photon interaction.  When $\omega$ is away from  $\omega_0$, for instance, when $\omega>1.01\omega_0$ for the ten-layer, $v_g$ can be approximated by LO phonon group velocity [Eq.~(\ref{lovg1app})], i.e. $v_g\propto \mathcal{N}$, whereas the PhP DOS is inversely proportional to $\mathcal{N}^2$, $g_p\propto 1/\mathcal{N}^2$ . Both simple relations on layer number  $\mathcal{N}$ have been verified and also demonstrated in Fig.~\ref{fig3}(a).  
While PhPs with a large $v_g$ in thick $\mathcal{N}$-layers are useful for waveguiding, the slow-light PhP modes in a thin layer can enhance light-matter interactions \cite{Basov:2016,Low:2017}.   

\subsection{PhPs in $\mathcal{N}$-layer hBN on SiO$_2$ and DE effects }

For $\mathcal{N}$-layer hBN on a SiO$_2$ substrate, the dielectric response of SiO$_2$   $\epsilon_1(\omega)$ should be considered to obtain the 2D hBN PhPs. We use a lattice DF accounting for two TO modes \cite{Fischetti:2001,Hauber:2017} for $\epsilon_1(\omega)$, 
\begin{equation}
\epsilon_1(\omega)=\epsilon_{1,\infty}+\frac{\epsilon_{1,0}-\epsilon_{1,i}}{1-(\omega/\omega_{T1})^2}+\frac{\epsilon_{1,i}-\epsilon_{1,\infty}}{1-(\omega/\omega_{T2})^2},    
\label{ldfsio2}
\end{equation}
where $\epsilon_{1,0}$ and $\epsilon_{1,\infty}$ are the static and high-frequency DCs,  $\epsilon_{1,i}$ is an intermediate DC from parameterization, and $\omega_{T1}$ and $\omega_{T2}$ are the two TO mode frequencies, $\omega_{T1}<\omega_{T2}$. From Refs.\cite{Fischetti:2001,Hauber:2017} 
$\epsilon_{1,\infty}$=2.5, $\epsilon_{1,i}$=3.05, $\epsilon_{1,0}$=3.9, $\omega_{T1}$=55.6 meV and $\omega_{T2}$=138.1 meV for SiO$_2$. Only 
when $\omega\gg\omega_{T2}$ can the ions in SiO$_2$ be clamped with $\epsilon_1(\omega)=\epsilon_{1,\infty}$.  Clearly    
$\omega$ in the PhP frequency band is not much higher than $\omega_{T2}$  (see Fig.~\ref{fig9} in Appendix A), thus 
$\epsilon_1(\omega)\neq\epsilon_{1,\infty}$ and more precisely $1.33\le\epsilon_1(\omega)\le1.92<\epsilon_{1,\infty}$ due to the ionic motions of SiO$_2$. Therefore the lattice vibrations in connection with the 2D PhPs occur in both $\mathcal{N}$-layer hBN and SiO$_2$. The calculated PhP dispersions  are shown in Fig.~\ref{fig2}(b).
Given frequency $\omega$ the confinement factor $\beta$ is larger than that of a freestanding $\mathcal{N}$-layer.   
This is because with a refractive index greater than 1, the substrate causes a decrease in the phase velocity and wavelength of the PhP waves.  At centre frequency $\omega_c\approx185$ meV, for instance, $\beta$ is increased by $\sim$36\% in ML and ten-layer hBN. Neglecting HFS, as the ten-layer on SiO$_2$ result (dotted line) shows, the underestimate of $\beta$ remains large, as in the freestanding layers above [Fig.~\ref{fig2}(a)]; e.g., there is a 49\% underestimate at $\omega_c$.  
The ESA again yields proper PhP dispersion for frequencies above $\omega_0$ (e.g., $\omega>1.01\omega_0$ for ten-layer hBN on SiO$_2$), but it becomes  invalid in the neighbourhood of $\omega_0$ [Fig.~\ref{fig2}(b) inset]. According to Eq.~(\ref{taylor3ap}), in the PhP dispersion near $\omega_0$ [solid curves in the inset of Fig.~\ref{fig2}(b)] there is mixing with the 3D PhPs of SiO$_2$ (shown as dashed line in the inset, which is not exactly vertical), due to the {\it coupled} lattice vibrations of hBN layers and SiO$_2$. As $\omega$ approaches $\omega_0$ the PhPs behave like 3D PhPs of SiO$_2$, with dispersion  $\beta^2=-\omega_0^3\epsilon_1^{\prime}(\omega_0)/(2\omega^2)+[\epsilon_1(\omega_0)+\omega_0\epsilon_1^{\prime}(\omega_0)/2]$ in the ($\omega$,$\beta$) plot (see Appendix A).

Dielectric effects of the substrate of 2D semiconductors are usually evaluated with a constant DC, e.g., the high-frequency or the static DC, in exciton \cite{Pedersen:2016} and phonon \cite{Sohier:2017} calculations. Here we calculated the PhP dispersion of ML and ten-layer hBN on SiO$_2$ using three DCs $\epsilon_{1,\infty}$, $\epsilon_{1,0}$, $\epsilon_1(\omega_0)$ of SiO$_2$, respectively, and then compare  with the results obtained with frequency-dependent DF $\epsilon_1(\omega)$  [Eq.~(\ref{ldfsio2})] in Fig.~\ref{fig4}. As $\epsilon_{1,\infty}$ has no contribution from ionic motion, whereas $\epsilon_{1,0}$ corresponds to ionic motion of SiO$_2$ at a very low frequency, both DCs result in a large discrepancy significantly overestimating confinement factor $\beta$ compared to the DF $\epsilon_1(\omega)$ calculation (solid curve);  e.g., using $\epsilon_{1,\infty}$ ($\epsilon_{1,0}$), $\beta$ is overestimated by 35\% (85\%) at $\omega$=180 meV for the ML. A larger DC corresponds to a greater refractive index and a smaller phase velocity, thus leading to a larger confinement factor. The dispersion curve calculated with $\epsilon_1(\omega_0)$, i.e. the response at minimum PhP frequency $\omega_0$, is closer to the $\epsilon_1(\omega)$ result, with $\beta$ being within a $\sim$14\% of deviation.

We now look at PhP group velocity $v_g$ and DOS $g_p$ for an $\mathcal{N}$-layer on SiO$_2$ with DF $\epsilon_1(\omega)$ [Eq.~(\ref{ldfsio2})]. 
As there is a several orders of magnitude fall of group velocity near $\omega_0$,  from the light speed to the LO mode group velocity, to calculate $v_g$ and the subsequent DOS accurately we use the analytical expression (\ref{povgdi2app}) in conjunction with the numerical solution of PhP modes. Compared to the freestanding $\mathcal{N}$-layer result [Fig.~\ref{fig3}(a)], clearly the PhP group velocity is decreased while the DOS is increased due to the substrate [Fig.~\ref{fig3}(b)]. 
Using a constant $\epsilon_1$ however makes $v_g$ and $g_p$ deviate significantly (not shown) from these calculated with DF $\epsilon_1(\omega)$. Using $\epsilon_1(\omega_0)$, for instance, the calculated $v_g$ is two times larger near $\omega_0$ and $\sim$20\% larger at the high frequencies. This is because apart from $\epsilon_1(\omega)$, the response change with respect to frequency $\epsilon^{\prime}_1(\omega)$ is also involved [see Eq.~(\ref{povgdi2app})].  
Away from $\omega_0$, $v_g$ and $g_p$ can be evaluated with simple ESA expressions [Eq.~(\ref{lovgdi2app}) for $v_g$].    


\subsection{PhP dispersion: comparison with experiment}


PhPs with confinement factor $\beta$ up to 300 were measured recently using EELS for {\it freestanding} hBN layers with different thicknesses (ML, 3, 4, 10 nm) \cite{Ningli:2020}. The measured PhPs have wavevectors $k\ge 12k_0$, thus far above the {\it wavevector gap}. We calculated PhPs in these hBN layers, i.e., ML  $\approx$0.32 nm, 9 layers (9L) $\approx$3 nm, 12 layers (12L) $\approx$4 nm and 31 layers (31L) $\approx$10 nm (interlayer distance is $c$/2, $c$ being the bulk hBN lattice parameter, $c$=6.425 $\AA$ \cite{Giles:2018}), and show the calculated and measured PhP dispersions $\omega$ versus $\beta$ in Fig.~\ref{fig5}(a) for a quantitative comparison (error bar length $\sim$3.75 meV is taken from Ref.\cite{Ningli:2020}). 
For the three MuLs (9L, 12L and 31L) the experimental data are quite close to the calculation, and in particular for 9L and 12L hBN the experimental and theoretical dispersion relations almost coincide for $\beta$ below 110.  Further, the calculated and measured low-frequency $v_g$ values ($\sim$$10^{-3}c$) are very close in the 9L and 12L.     
For the ML 
the measured PhP frequencies are approximately constant, very different from the theoretical result, i.e., an approximately linear increase of frequency with $\beta$ (bottom line).  
For the experimental $\beta$ values ($1/\beta$ is small, $1/\beta<0.028$ and $\omega>1.01\omega_0$), the ESA is a good approximation so the PhPs in the ML are like LO phonons [refer to Fig.~\ref{fig2}(a) above]. Indeed the calculated PhP group velocity $v_g\approx0.0001c$ [Fig.~\ref{fig3}(a)] is very close to the LO mode group velocities calculated from first principles \cite{Sohier:2017,Ferrabone:2011,Erba:2013} (see Ref.\cite{Zhang:2020a}). This $v_g$ is one order of magnitude greater than the experimental value $\sim$$10^{-5}c$ of ML hBN \cite{Ningli:2020} as the measured flat dispersion curve gives a very small slope. This suggests that to obtain accurate PhP dispersion more EELS experiment needs to be performed on ML samples, or in combination with an optical s-SNOM measurement. Neglecting HFS causes a large discrepancy between the calculation and experiment [as shown with 12L hBN, dashed line in Fig.~\ref{fig5}(a)].

For all four layers, the calculated PhP frequencies at the small wavevectors (low $\beta$) are very close to the measurements. On the large $k$ (or $\beta$) side the experimental frequency values are smaller, the deviation being within 2.0\%, 1.7\%, 0.5\% and 1.3\% for the ML, 9L, 12L and 31L, respectively. Phonon damping is ruled out as being responsible as the calculated result stays almost invariable with damping rate up to a large number $\gamma$=30 cm$^{-1}$.  In EELS the constructive interference between the excited and edge-reflected PhPs is used. For large-momentum ($\hbar k$) measurements   
the electron-beam probe is near to the edge of the sample (e.g., the distance is $\sim$10 nm for those data of $\beta\sim$ 300) and a large error may arise due to imperfections in the edge \cite{Ningli:2020} apart from the minimum energy resolution of 7.5 meV. 
Nevertheless, overall the calculated dispersion relations are in good agreement with  experiment for the three MuLs.

We now turn to PhPs in hBN MLs and bilayers on a SiO$_2$ substrate [Fig.~\ref{fig5}(b)]. 
The experimental dispersion relations were measured using s-SNOM \cite{Dai:2019}, with a minimum momentum compensation 11.75$k_0$ very close to that of the EELS measurements \cite{Ningli:2020} and a maximum momentum compensation $5.4\times 10^5/cm$ (according to $k/k_0\leq 61.7$  \cite{Dai:2019})  which is 18\% the maximum in the EELS experiment  (Fig.~4a of Ref.\cite{Ningli:2020}). For quantitative comparison a smaller frequency range 169 meV $\leq\omega\leq$ 173.5 meV than in Fig. 3 of  Ref.\cite{Dai:2019} (169 meV$\leq\omega\leq 176$ meV) is used for the  vertical axis. 
The PhP dispersions of ML (circles) and bilayer hBN on SiO$_2$ (triangles) are approximately linear through the measured confinement factor region $12<\beta<62$ (error bars are taken from Ref.\cite{Dai:2019}).   
The experimental data and theoretical result are quite close for the bilayer  
 apart from the lowest frequency datum (170 meV).   
For the ML   
the experimental data of PhP frequencies below 171 meV are near the calculated dispersion curve (lower solid line), 
while the two large-$\omega$ data show a significant deviation from the calculation. At $\omega$=171.7 meV, for instance, the measured $\beta$ value  is 41\% smaller. In fact, a larger error occurred in the larger-momentum measurements \cite{Dai:2019,Shiz:2015}, and this measurement error reached 50\% for the 171.7 meV datum of the ML. The experimental group velocity $0.0002c$ of the ML, one order of magnitude larger than that measured with EELS \cite{Ningli:2020} of the freestanding ML [Fig.~\ref{fig5}(a)], is closer to the calculated value $\sim$$0.0001c$ for both the freestanding and supported ML. Considering these, nonetheless, good agreement is obtained between the calculated and experimental results for both supported layers.   
Furthermore,   
the measured large-momentum PhPs are closer to the dispersion curves of the freestanding ML and bilayer (dashed lines), suggesting a weaker screening effect from SiO$_2$.  As the momentum increases, more electric field of the lattice polarization charges is confined within the 2D materials, leading to the weakened screening from the substrate \cite{Sohier:2017}.


\subsection{Field confinement, PQF and wavelength scaling}

The electric and magnetic fields of the PhPs decay exponentially according to $e^{-K_1\lvert z\rvert}$ and $e^{-K_2\lvert z\rvert}$ in the two half-space media respectively.  
Thus, to quantify the confinement to the interface of the PhP waves, $z_{c,1}=1/K_1$ and $z_{c,2}=1/K_2$ are used to define the field decay lengths of the evanescent mode in the $z$ direction \cite{Maier:2007}. Our numerical calculation indicates that for on-SiO$_2$ hBN layers the decay lengths in air and the substrate are nearly equal, $z_{c,1}\approx z_{c,2}$ (thus a simple $z_c$ can be used), when frequencies $\omega>1.01\omega_0$ approximately.  As shown in Fig.~\ref{fig6}(a),  a 
higher-frequency PhP wave has a shorter decay length $z_c$ and thus greater confinement to the interface. As a higher frequency $\omega$ corresponds to a smaller vacuum wavelength  $\lambda_a$ (upper horizontal axis), PhPs excited with shorter wavelength infrared light are more confined to the interface. A decay length below 1 nm can be reached with an excitation wavelength shorter than 6.4 nm for freestanding 5L hBN, for instance. A thinner hBN $\mathcal{N}$-layer causes larger confinement to the PhPs and the substrate introduces further confinement.  
We also calculated the PhP propagation length $L_p=1/\alpha_{\mathcal{N}}(\omega)$, $\alpha_{\mathcal{N}}(\omega)$ being the $\mathcal{N}$-layer absorption coefficient (section III A), and show the results in Fig.~\ref{fig6}(b). In the $\mathcal{N}$-layers on SiO$_2$ $L_p$ is lower by a factor of $(\epsilon_1(\omega)+1)/2$ compared with that of the freestanding layers [refer to Eq.~(\ref{lonlay0})]. Evidently there is a fundamental trade-off between loss and localization like in surface plasmon polaritons \cite{Maier:2007}, i.e.,   
PhPs in a thinner $\mathcal{N}$-layer or a stronger DE exhibit greater confinement and a shorter propagation distance.  


The PQF $\gamma_p^{-1}$ is calculated from Eq.~(\ref{lonlay0}) giving  $\gamma_p^{-1}$=$\lvert$ Re($\chi(\omega)$)/Im($\chi(\omega)$)$\rvert$. That is, the PQF is independent of the number of layers, 
$\gamma_p^{-1}=-[(\omega/\omega_0)^3-(\eta+2-\gamma^2/\omega_0^2)\omega/\omega_0+(1+\eta)\omega_0/\omega]\omega_0/\eta\gamma$.  
Fig.~\ref{fig7} shows the PQFs versus PhP frequency $\omega$ (solid lines) for 
two damping rates $\gamma=$7 and 15 cm$^{-1}$. The PQFs at the two bounds $\omega_0$ and $\omega_u$, i.e. $\gamma/(\eta\omega_0)$ and $\gamma\sqrt{1+\eta}/(\eta\omega_0)$, are both very small as $\gamma\ll\omega_0$. The maximum PQF $\gamma_{p,m}^{-1}$ occurs at $\omega_{p,m}^2=\omega_0^2(\eta+2+\sqrt{\eta^2+16\eta+16})/6$.  Clearly, 
$\omega_{p,m}=\omega_c=\omega_0$, if $\eta=0$. It is found by a simple numerical calculation that 
 $\omega_{p,m}$ is almost equal to $\omega_c$ for $0<\eta<3$ (e.g., at $\eta$=2, $\omega_{p,m}=1.36693\omega_0$, $\omega_c=1.36602\omega_0$), and as $\eta$ increases their difference becomes larger (e.g. at $\eta$=5, $\omega_{p,m}=1.73205\omega_0$, $\omega_c=1.72474\omega_0$). For hBN layers $\eta=0.38052$, $\omega_{p,m}=1.087484\omega_0=184.853$ meV, $\omega_c=1.087478\omega_0$; therefore the PQF peaks almost at the centre frequency. 
 The maximum PQF decreases from 17.09 to 7.95 as the damping rate $\gamma$ increases from 7 cm$^{-1}$ to 15 cm$^{-1}$.  
PhPs with a large PQF are useful for practical applications involving PhP propagation \cite{Basov:2016,Low:2017}. Our result indicates that the frequencies of such polaritons should be near  the center frequency;  in particular these frequencies should stay away from $\omega_u$, the upper frequency bound, where the minimum PQF (Fig.~\ref{fig7}) and propagation length [Fig.~\ref{fig6}(b)] both occur. This is distinct from the previous finding obtained with no HFS \cite{Rivera:2019}:   
 neglecting HFS yields PQF  $\gamma_p^{-1}=[\omega/\omega_0-\omega_0/\omega]\omega_0/\gamma$, which  increases with the frequency monotonically across the PhP frequency band and also beyond [dotted lines in Fig.~\ref{fig7}, similar to the Fig.~2 (b) results of Ref.\cite{Rivera:2019}]. Now the PQF  $(\sqrt{1+\eta}-1/\sqrt{1+\eta})\omega_0/\gamma$ at $\omega_u$ is much higher than the PQF peak with the nonlocal HFS included. With $\gamma$=7 cm$^{-1}$, for instance, neglecting HFS  overestimates the maximum PQF by a factor of 2.7.

To see the effect of dimensionality change we also calculated the PQFs of bulk hBN on SiO$_2$ using Eq.~(1) of Ref.\cite{Dai:2014} for the two damping rates (dashed curves; $\gamma=$7 cm$^{-1}$ is an experimental value from Refs.\cite{Giles:2018,Caldwell:2014}). Including the HFS, the 2D PhPs have almost the same band of frequencies as the 3D PhPs. At a low frequency $\omega<177$ meV the 2D and 3D PhPs have the same PQF approximately, whereas at the higher frequencies there is a lower PQF for the 2D PhPs ($\sim$8\% lower than the PQF of 3D PhPs at $\omega_c$). No experimental PQFs are found for hBN layers but we add the experimental data of bulk hBN on SiO$_2$ from Ref.\cite{Giles:2018} (circles) and Ref.\cite{Dai:2014} (square) to Fig.~\ref{fig7}, which are close to the calculation (upper dashed curve, $\gamma=$7 cm$^{-1}$).

Given an incident frequency $\omega$, scaling of the PhP wavelength $\lambda$ with material thickness $d$ is a key PhP property  \cite{Basov:2016,Dai:2014,Shiz:2015,Dubrovkin:2018,Dai:2019}. The experimental study with hBN on SiO$_2$ \cite{Dai:2014} found that for a {\it high} frequency $\omega=1.14\omega_0$=1560 cm$^{-1}$ $\lambda$ increases linearly with $d$ in bulk hBN (Fig.~3E of Ref.\cite{Dai:2014}; 83 nm $\le d \le$ 445 nm), while at a lower frequency  slightly above $\omega_0$ ($\omega=1.02\omega_0$=1400 cm$^{-1}$) the PhP scaling becomes nonlinear in {\it thin} layers (4 nm $\le d \le$ 53 nm), with a smaller slope at larger thickness $d$ (Fig.~3E inset of Ref.\cite{Dai:2014}). Similar linear and nonliear scaling results were observed also in another experimental study for SiO$_2$-supported hBN \cite{Shiz:2015} (Fig.~6C therein).  
 In hBN layers, 
{\it within the ESA} the wavelength is proportional to the number of layers $\mathcal{N}$ from Eq.~(\ref{lonlay0}),  $\lambda=-8N\pi^2\chi(\omega)/(\epsilon_1(\omega)+\epsilon_2(\omega))$.  Indeed, the numerical solution of Eq.~(\ref{polaton0}) indicates that this scaling law is valid only for frequencies away from $\omega_0$, e.g., $\omega >1.07 \omega_0$ approximately for $d<$ 35 nm, which is demonstrated in Fig.~\ref{fig8}(a) with two incident frequencies  $\omega$=193.4 meV, i.e. 1560 cm$^{-1}$ in experiment \cite{Dai:2014} and $\omega$=182.3 meV, i.e. 1470 cm$^{-1}$ in experiment \cite{Shiz:2015}, for both freestanding (dashed lines) and on-SiO$_2$ (solid lines) hBN layers. At both frequencies the phonon contribution to the PhPs is dominant (refer to Figs.~\ref{fig2} and \ref{fig3}; note that the incident $\omega$=182.3 meV in experiment \cite{Shiz:2015} is near centre frequency $\omega_c$=184.9 meV). We calculated PhP wavelength scaling for bulk hBN using Eq.~(1) of Ref.\cite{Dai:2014} 
and show the results for both frequencies in Fig.~\ref{fig8}(b) together with the obtained thin-layer results  
[dashed lines: freestanding hBN; solid lines: on-SiO$_2$ hBN; for parameters of bulk calculations see Ref.\cite{Dai:2014} Supplementary Materials]. 
The 2D and 3D PhP scaling are both linear as both 2D and 3D PhPs are phonon-like. 
Similar results are found also for other frequencies when the phonon content in PhPs is very large.    
We also add the experimental data of on-SiO$_2$ hBN layers \cite{Shiz:2015} and bulk hBN \cite{Dai:2014,Shiz:2015} to Fig.~\ref{fig8}(a) and (b) (circles and diamonds) and find good agreement between theory and experiment (in thin layers the experimental data are closer to the freestanding-layers calculation).

 At lower incident frequencies  $\omega$=173.6 and 174.8 meV, corresponding to 1400 and 1410 cm$^{-1}$, respectively, used in experiments \cite{Dai:2014,Shiz:2015}, the wavelength scaling is calculated with Eq.~(\ref{polaton0}) and   
deviation from the linear law occurs when the layer thickness $d$ exceeds $\sim$15 nm [lower two pairs of curves in Fig.~\ref{fig8}(c); dashed for freestanding hBN and solid for on-SiO$_2$ hBN] due to the phonon-photon coupling in the PhPs. These calculated results for hBN on SiO$_2$ are in good agreement with the experimental data  \cite{Dai:2014,Shiz:2015} [squares and triangles in Fig.~\ref{fig8}(c)]. 
The nonlinear scaling becomes pronounced as $\omega$ decreases, as shown with $\omega$=171.7 meV, because the phonon-photon interaction is enhanced (Figs.~\ref{fig2} and \ref{fig3}).  
 Further, the PhP scaling exhibits saturation for $\omega$ near $\omega_0$; for instance, at $\omega$=170.2 meV $\lambda$ stays nearly constant ($\approx$ photon wavelength at $\omega_0$), i.e. independent of the layer thickness when $d>$ 20 nm [top two lines in Fig.~\ref{fig8}(c)], which can be well described by the expressions in Appendix A [e.g. Eq.~(\ref{scalapp}) for freestanding layers].  Physically these PhPs in the freestanding (supported) layers with $d>$ 20 nm are free-space photons (PhPs of the substrate) approximately with frequency $\omega_0$, and therefore the wavelength does not vary with $d$ (Appendix A).

\section{Conclusions}


We have studied the in-phase PhPs in both freestanding and on-substrate hBN MuLs using the macroscopic optical-phonon model.  
The PhP modes are derived from Maxwell's equations by considering the $\mathcal{N}$-layer as a 2D dielectric embedded between two half-space crystals. 
The PhP confinement, group velocity, PQF and wavelength scaling are calculated. The  
effects of the nonlocal HFS on the key PhP properties are studied and the ESA is examined.   Owing to the HFS, 
the upper bound of PhP frequency occurs, limiting the PhP frequencies in a finite range, and also the PQF varies with frequency $\omega$ similarly to that of bulk hBN, having the maximum near centre frequency $\omega_c$ which is slightly ($\sim$8\%) smaller than the bulk's maximum PQF.     
 Including the HFS also significantly increases the confinement factor (e.g. by 48\% at $\omega_c$). 
The RE is so strong near TO mode frequency $\omega_0$ that there is a wavevector gap with no PhP modes in it ($k<\omega_0/c$). 
The ESA is an effective approach for $\omega$ above $\omega_0$ (e.g. $\omega>1.03\omega_0$ for 30-layers) but it fails to describe the PhP properties near $\omega_0$ due to the strong RE.  At high frequencies when the phonon contribution is dominant, the PhP wavelength increases linearly with the MuL thickness, well describable with the ESA; at a lower frequency, however the PhP scaling deviates from the linear law owing to the phonon-photon coupling. In both respects the calculated results are in good agreement with experiment. Further at a frequency near $\omega_0$ the scaling saturation occurs as the PhPs are like free-space photons for freestanding MuLs and the substrate's 3D PhPs for supported layers. The calculated PhP dispersions are compared with the EELS data for freestanding hBN layers and with the s-SNOM measurements for ML and bilayer hBN on SiO$_2$, and good agreement between calculation and experiment is obtained except for the freestanding ML.    
The DE effects are discussed and the substrate's DF should be used to accurately calculate the PhP properties.

The incident light frequency should be near the centre frequency to maximize the PQF   
whereas the PhP wavelength, confinement and propagation distance are tunable by varying the MuL thickness and DE. Therefore, the desired PhP properties can be engineered for waveguiding applications.   
Our model can be extended to studying PhPs in MuLs of transition metal dichalcogenides such as MoS$_2$.



\begin{acknowledgments}
We acknowledge support from the New Energy and Materials Collaboration project of the School of Physics, and the Natural Science Research Funds (Nos. 419080500175 \& 419080500260) of Jilin University. 
\end{acknowledgments}

\newpage

\appendix

\section{PhP dispersion and scaling near TO phonon frequency in an $\mathcal{N}$-layer}   

 For a freestanding $\mathcal{N}$-layer, the PhP dispersion equation (\ref{polaton0}) can be nondimensionalized and transformed to 
\begin{equation}
x^2=y^2+\frac{1}{(2\pi \mathcal{N}\alpha)^2}\Big(\frac{\eta}{1+\eta-y^2}-1\Big)^2,       
\label{pofre5ap}
\end{equation}
where $1<y<\sqrt{1+\eta}$ (see section II C for the dimensionless quantities).  
Evidently $x^2$ increases monotonically with $y^2$ and $x>1$; thus $k_0$ is the {\it minimum} wavevector, and there are no PhPs when $k<k_0$. When $\omega$ is near $\omega_0$, by Taylor expansion to second order about $y^2=1$, one finds 
\begin{equation}
x^2=y^2+\Big(\frac{c}{2\mathcal{N}v_l}\Big)^2(y^2-1)^2,    
\label{taylor2ap}
\end{equation}
from which confinement factor $\beta$ can be expressed in terms of $\omega$ to give the PhP dispersion, 
\begin{equation}
\beta^2=1+\left[\frac{c}{2\mathcal{N}v_l}\Big(\frac{\omega}{\omega_0}-\frac{\omega_0}{\omega}\Big)\right]^2.    
\label{dispbeta}
\end{equation}

Note that neglecting HFS reduces Eq.~(\ref{pofre5ap}) straightforward to Eq.~(\ref{taylor2ap}). 

Relating wavevectors $k$, $k_0$ to wavelengths $\lambda$, $\lambda_0$, one obtains from Eq.~(\ref{taylor2ap}) the scaling of wavelength for frequencies near $\omega_0$,
\begin{equation}
\lambda=\frac{\lambda_0}{\sqrt{(\frac{\omega}{\omega_0})^2+\big\{\frac{c}{2\mathcal{N}v_l}[(\frac{\omega}{\omega_0})^2-1]\big\}^2}}~,   
\label{scalapp}
\end{equation}
which is demonstrated in Fig.~\ref{fig8}(c) (top dashed line).     

For an $\mathcal{N}$-layer of hBN in dielectrics with DCs $\epsilon_1$ (e.g. substrate) and $\epsilon_2$ (e.g. air) the PhP dispersion Eq.~(\ref{polaton0}) becomes   
\begin{equation}
\frac{\epsilon_1}{\sqrt{x^2-\epsilon_1
		y^2}}+\frac{\epsilon_2}{\sqrt{x^2-\epsilon_2
		y^2}}=4\pi \mathcal{N}\alpha\Big(\frac{\eta}{y^2-1}-1\Big).       
\label{pofre2ap}
\end{equation} 
Let $\epsilon_1>\epsilon_2$. When $\epsilon_1$ and $\epsilon_2$ are 
taken to be independent of frequency $\omega$,  $x>\sqrt{\epsilon_1}$; i.e., the minimum wavevector is $\sqrt{\epsilon_1}k_0$, larger than that of freestanding layers.  Similarly, expanding function $x^2$ to second order in $y^2-1$, the PhP dispersion near $\omega_0$ is given by 
\begin{equation}
\beta^2=\epsilon_1+\left[\frac{\epsilon_1c}{4\mathcal{N}v_l}\Big(\frac{\omega}{\omega_0}-\frac{\omega_0}{\omega}\Big)\right]^2,  
\label{dispbetadia}
\end{equation}
and the wavelength scaling near $\omega_0$ is 
\begin{equation}
\lambda=\frac{\lambda_0}{\sqrt{\epsilon_1(\frac{\omega}{\omega_0})^2+\big\{\frac{\epsilon_1c}{4\mathcal{N}v_l}[(\frac{\omega}{\omega_0})^2-1]\big\}^2}}~.  
\label{scalappdia}
\end{equation}

In fact the substrate's DC $\epsilon_1$ depends on $\omega$ due to its lattice vibrations [$\epsilon_1(\omega)$ of SiO$_2$ (Eq.~(\ref{ldfsio2})) is shown in Fig.~\ref{fig9}]. $\epsilon_1$ can be readily changed to a function of $y^2$, denoted as $\bar{\epsilon}_1(y^2)$; i.e.,   $\bar{\epsilon}_1(y^2)=\epsilon_1(\omega)$.  Substituting $\bar{\epsilon}_1(y^2)$ for $\epsilon_1$ in Eq.~(\ref{pofre2ap}), a lengthy derivation yields a Taylor expansion of function $x^2$ about $y^2=1$, independent of the form of $\epsilon_1(\omega)$, 
\begin{align}
x^2&=-\bar{\epsilon}^{\prime}_1(1)+[\bar{\epsilon}_1(1)+\bar{\epsilon}^{\prime}_1(1)]y^2
\nonumber \\ &\qquad {} 
+\Big[\Big(\frac{\bar{\epsilon}_1(1)c}{4\mathcal{N}v_l}\Big)^2+\bar{\epsilon}^{\prime}_1(1)+\frac{1}{2}\bar{\epsilon}^{\prime\prime}_1(1)\Big](y^2-1)^2,    
\label{taylor3ap}
\end{align}
involving derivatives at $y^2=1$, i.e. $\bar{\epsilon}^{\prime}_1(1)$ and    $\bar{\epsilon}^{\prime\prime}_1(1)$; note that $\bar{\epsilon}^{\prime}_1(y^2)$ and  $\bar{\epsilon}^{\prime\prime}_1(y^2)$ are derivatives with respect to $y^2$, 
$\bar{\epsilon}^{\prime}_1(y^2)=d\bar{\epsilon}_1(y^2)/dy^2$, and $\bar{\epsilon}^{\prime\prime}_1(y^2)=d\bar{\epsilon}^{\prime}_1(y^2)/dy^2$. 
When 3D PhP dispersion of the substrate $c^2k^2/\omega^2=\epsilon_1(\omega)$ \cite{Born:1954}, i.e., $x^2=y^2\bar{\epsilon}_1(y^2)$ after nondimensionalization, is expanded about $y^2=1$ to second order as
$x^2=-\bar{\epsilon}^{\prime}_1(1)+[\bar{\epsilon}_1(1)+\bar{\epsilon}^{\prime}_1(1)]y^2+[\bar{\epsilon}^{\prime}_1(1)+\frac{1}{2}\bar{\epsilon}^{\prime\prime}_1(1)](y^2-1)^2$,    
one immediately finds that all these terms appear exactly in expansion (\ref{taylor3ap}) above. The substrate's 3D PhPs are involved as both lattice vibrations of $\mathcal{N}$-layer hBN and the substrate are coupled with the electromagnetic field. The scaling of wavelength can be obtained straightforward from Eq.~(\ref{taylor3ap}) [refer to the derivation of Eq.~(\ref{scalapp}) from Eq.~(\ref{taylor2ap})], which is demonstrated in Fig.~\ref{fig8}(c) (top solid curve).

For $\omega$ further close to $\omega_0$ ($y$ near 1) one can neglect the second order terms of the expansions. Then one has $x=y$, i.e., $\omega=ck$, and $\beta=1$ for freestanding $\mathcal{N}$-layers, and $x=\sqrt{\epsilon_1}y$, i.e., $\omega=ck/\sqrt{\epsilon_1}$, and $\beta=\sqrt{\epsilon_1}$ for $\mathcal{N}$-layer hBN on a substrate with constant $\epsilon_1$. Therefore the PhPs of the $\mathcal{N}$-layer behave like the optical waves (photons) as $\omega\rightarrow\omega_0$, and they also have the same phase velocity and group velocity, independent of the layer number $\mathcal{N}$. There is no scaling of  wavelength $\lambda$ with $\mathcal{N}$. When the substrate's dielectric response $\epsilon_1(\omega)$ is accounted for, the 2D PhPs behave like the PhPs of the substrate with dispersion $c^2k^2/\omega^2=-\omega_0^3\epsilon_1^{\prime}(\omega_0)/(2\omega^2)+[\epsilon_1(\omega_0)+\omega_0\epsilon_1^{\prime}(\omega_0)/2]$ from Eq.~(\ref{taylor3ap}). As $\epsilon_1^{\prime}(\omega_0)\ne 0$ [$\omega_0\epsilon_1^{\prime}(\omega_0)$=6.51 and $\epsilon_1(\omega_0)$=1.3 for SiO$_2$] the dispersion near $\omega_0$ is not linear  anymore, different from the constant $\epsilon_1$ case above. The confinement factor $\beta=\sqrt{\epsilon_1(\omega_0)}$ at $\omega_0$.

\section{PhP group velocity and density of states in an $\mathcal{N}$-layer}   

\begin{figure}
	\includegraphics*[width=8cm]{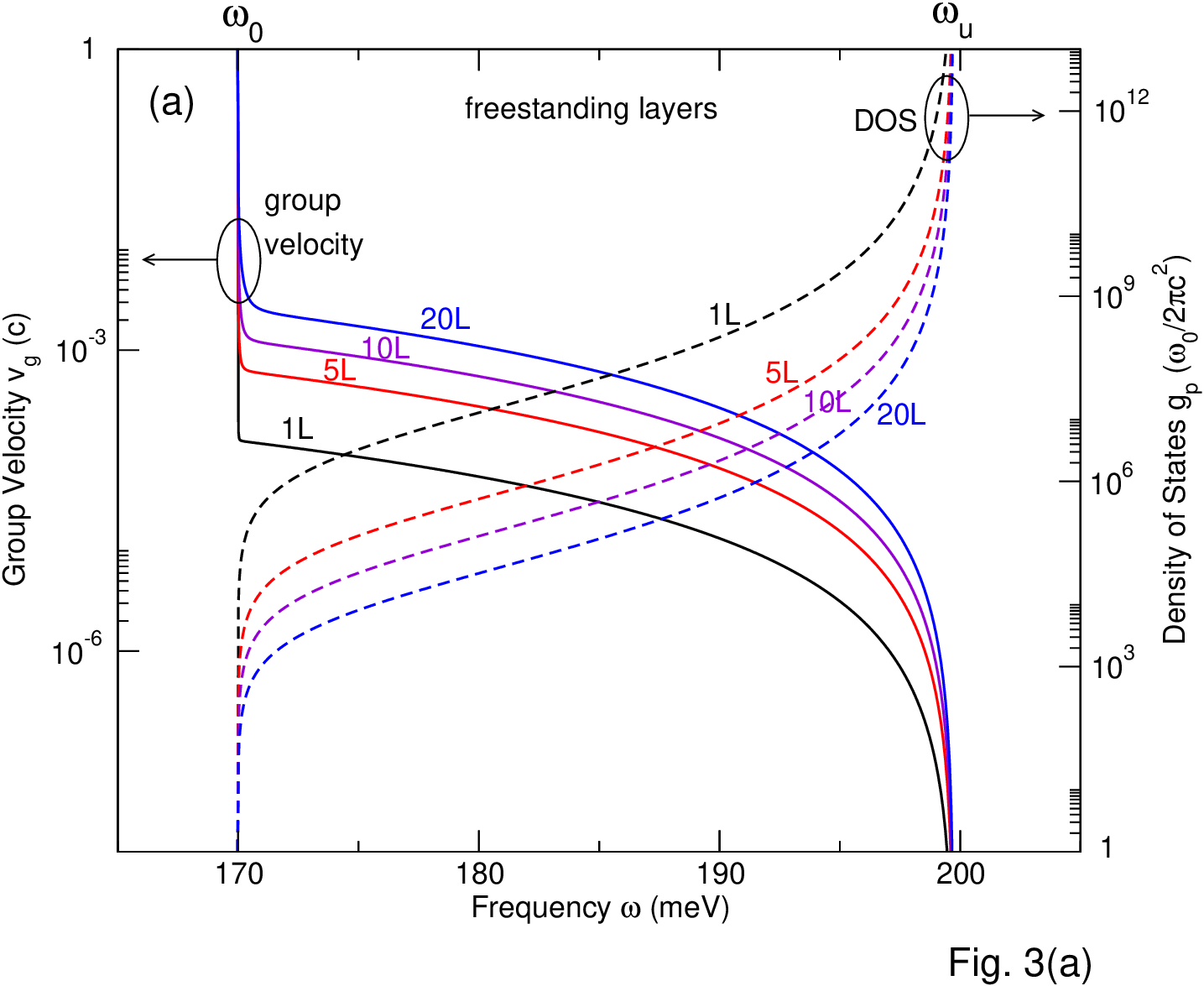}
	\includegraphics*[width=8cm]{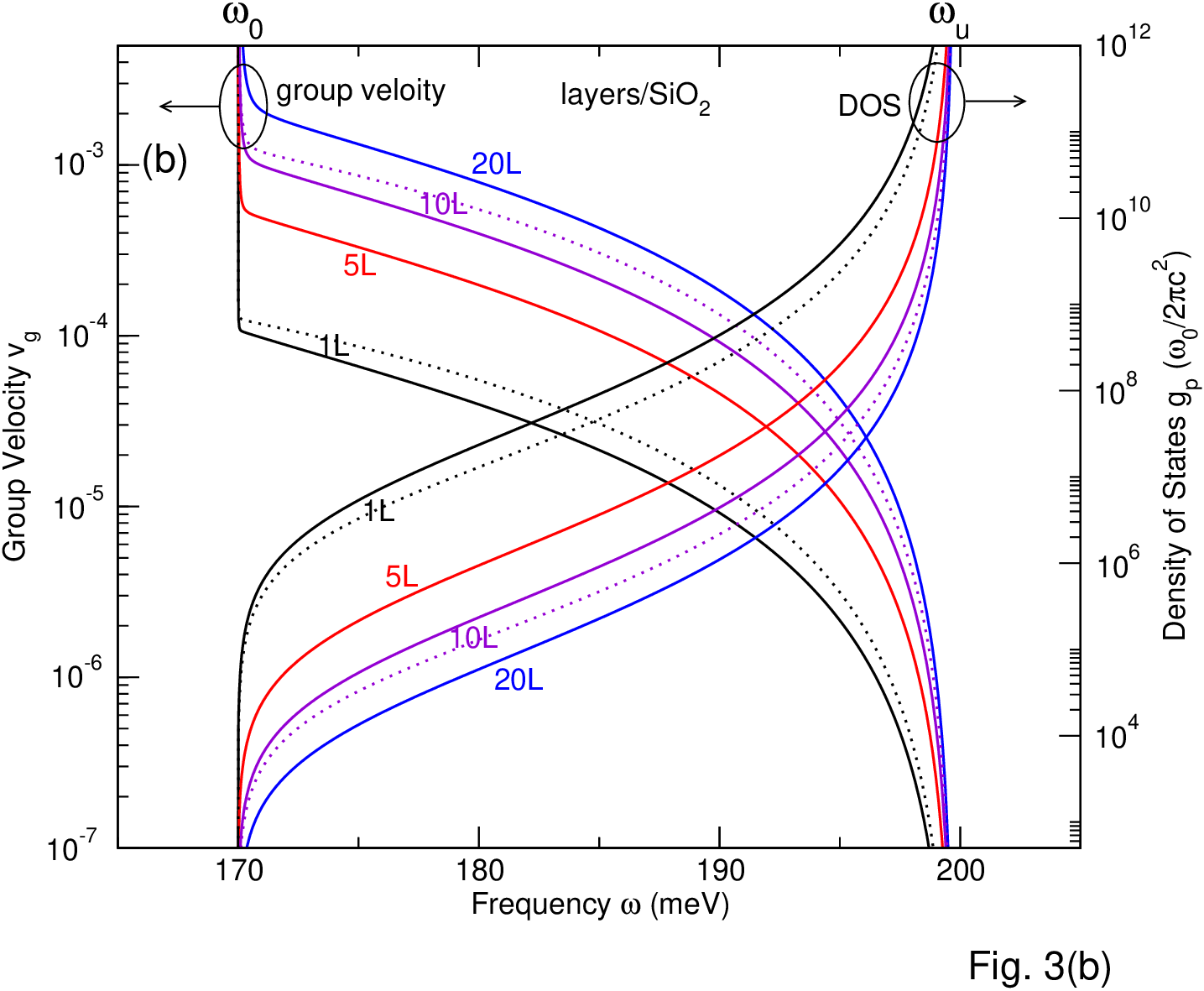}
	\caption
	{(Color online) PhP group velocities (in units of $c$; left vertical axis) and densities of states (in units of $\omega_0/2\pi c^2$; right vertical axis) versus frequency for (a) four freestanding hBN layers, i.e., an ML (1L), 5 layers (5L), 10 layers (10L) and 20 layers (20L) of hBN, and (b) the four hBN layers on SiO$_2$. The dielectric response $\epsilon_1(\omega)$ of SiO$_2$ [Eq.~(\ref{ldfsio2})] is included in (b), and the freestanding 1L and 10L results are also added to (b) (dotted  lines). The lower and upper PhP frequency limits $\omega_0$ and $\omega_u$ are labelled. 
	}
	\label{fig3}
\end{figure}

\begin{figure}
	\includegraphics*[width=10cm]{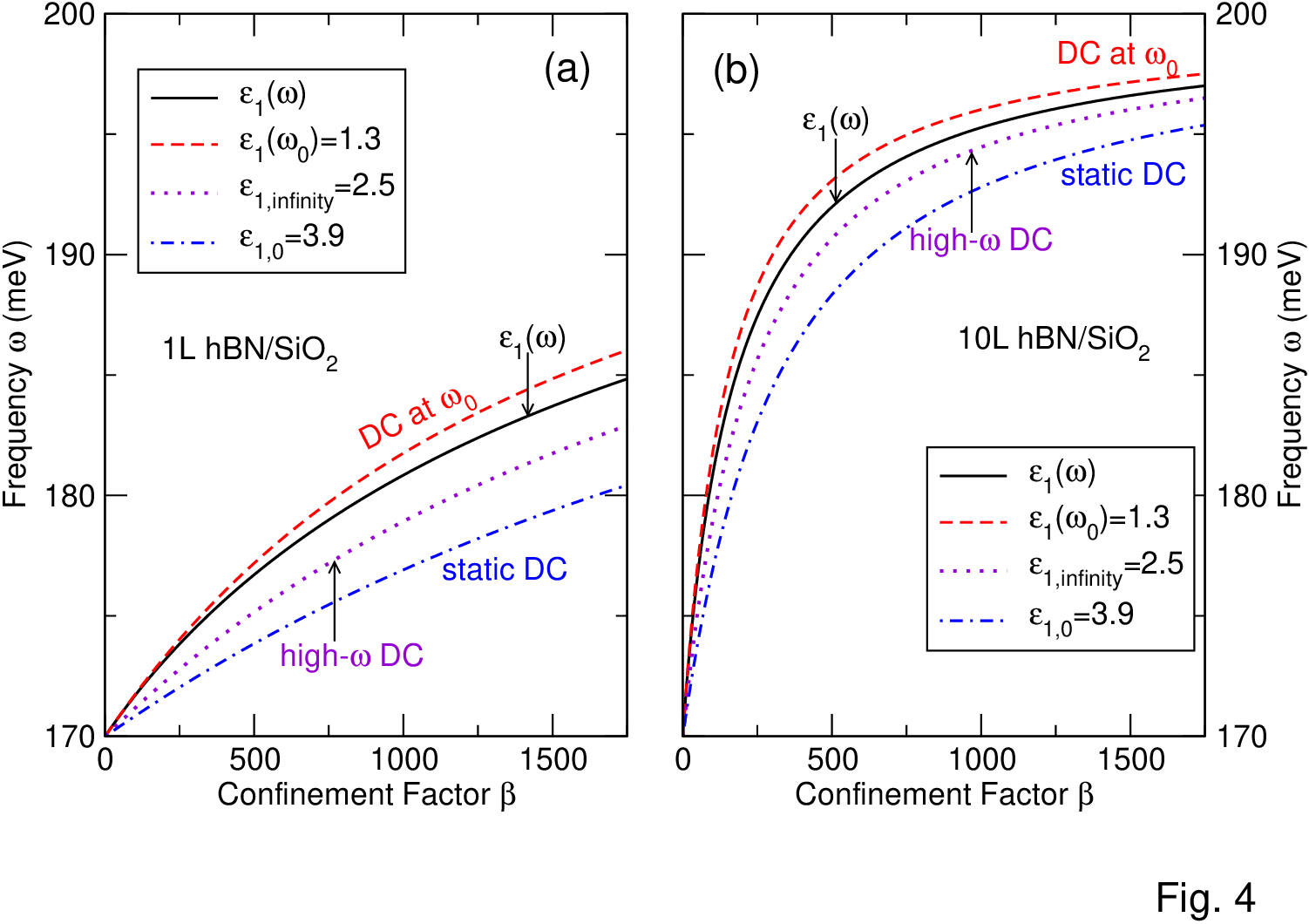}
	\caption
	{(Color online) Dispersion relations of PhPs in (a) ML and (b) ten-layer hBN on SiO$_2$, calculated using the DF of SiO$_2$ $\epsilon_1(\omega)$ [Eq.~(\ref{ldfsio2})] and also the three DCs of SiO$_2$, the high-frequency DC $\epsilon_{1,\infty}=2.5$, the static DC $\epsilon_{1,0}=3.9$ and the DC at the minimum PhP frequency $\epsilon_1(\omega_0)=1.3$, respectively. 
	}
	\label{fig4}
\end{figure}

\begin{figure}
	\includegraphics*[width=9cm]{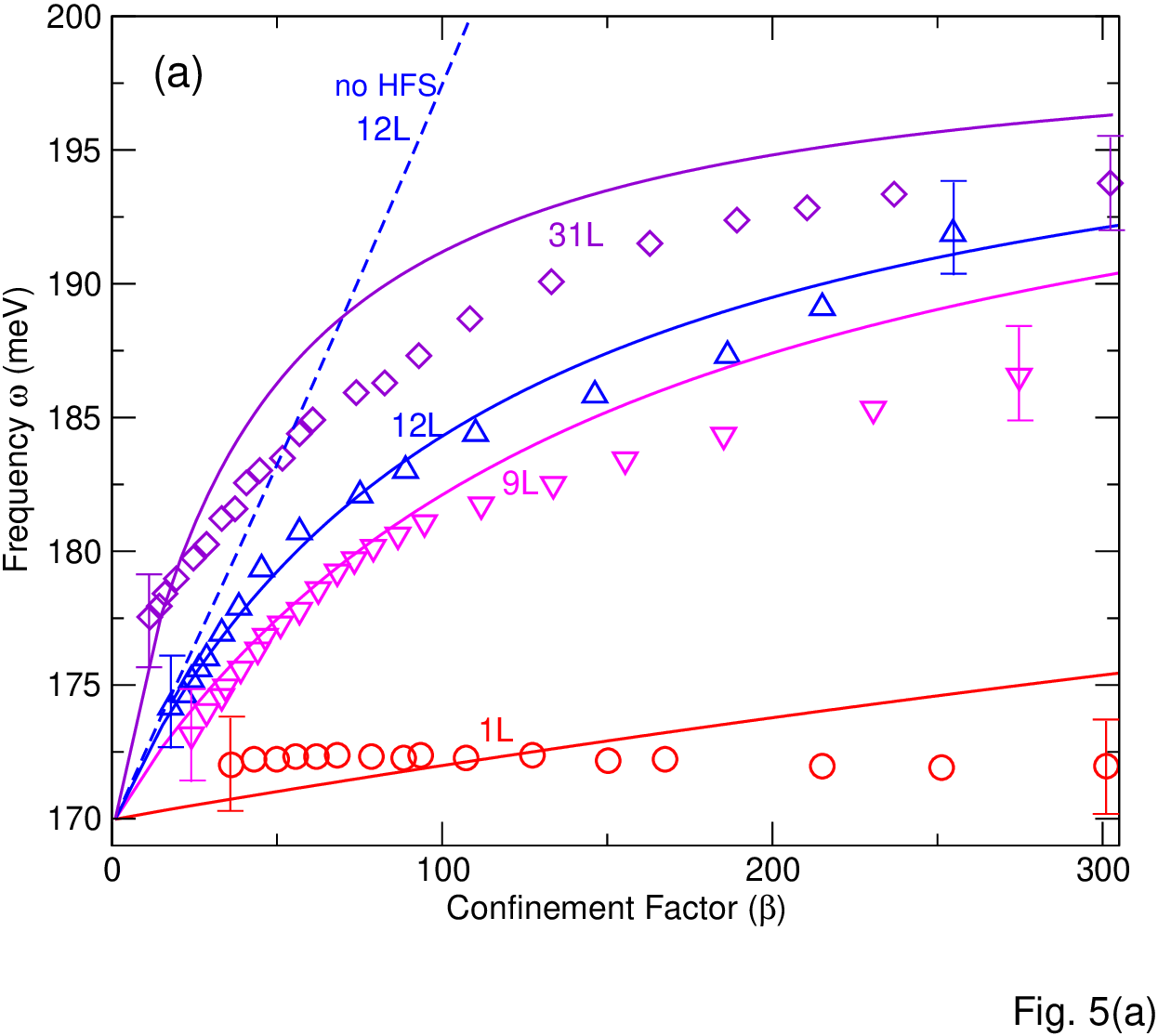}
	\includegraphics*[width=9cm]{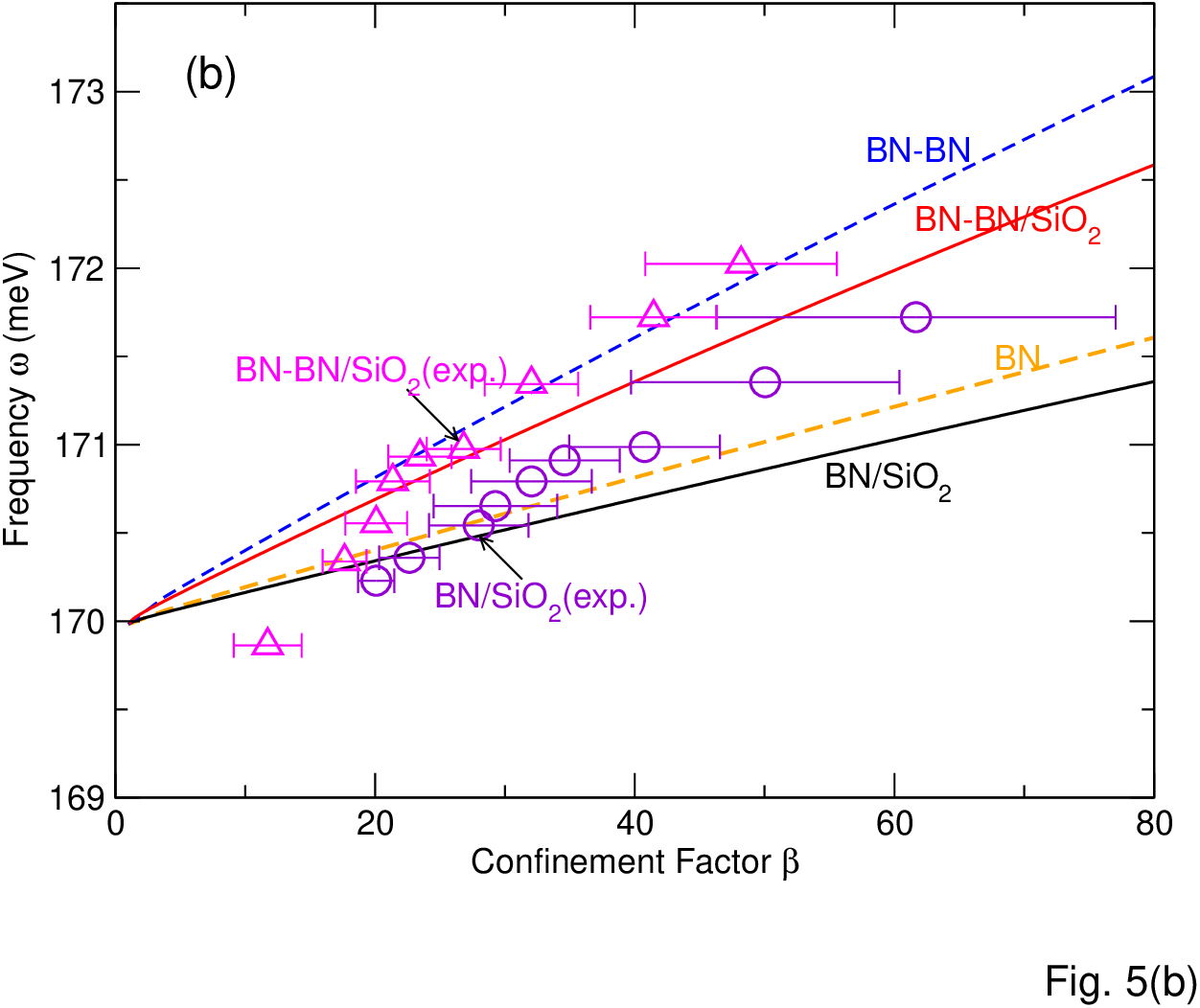}
	\caption
	{(Color online) Dispersion relations of PhPs in (a) four freestanding hBN layers with different thicknesses, i.e., ML (1L), 9 layers (9L), 12 layers (12L)  and 31 layers (31L), calculated from Eq.~(\ref{pofre5ap}) (solid curves) and measured with EELS (symbols, from Fig.~4a of  Ref.\cite{Ningli:2020}; measurement errors $\sim$3.75 meV are drawn for endpoint data), and (b) ML and bilayer hBN on SiO$_2$ calculated from Eq.~(\ref{pofre2ap}) with the DF of SiO$_2$ Eq.~(\ref{ldfsio2}) (solid lines) and measured using s-SNOM (symbols with an error bar, from Fig. 3 of Ref.\cite{Dai:2019}). 
		The calculated result without HFS is plotted for 12L hBN in (a) (dashed line), and the calculated freestanding ML and bilayer hBN results are also shown in (b) (dashed lines). The frequency range ($y$-axis) in (b) is 2.5 meV smaller than that of Fig. 3 of Ref.\cite{Dai:2019}.  
	}
	\label{fig5}
\end{figure}

\begin{figure}
	\includegraphics*[width=10cm]{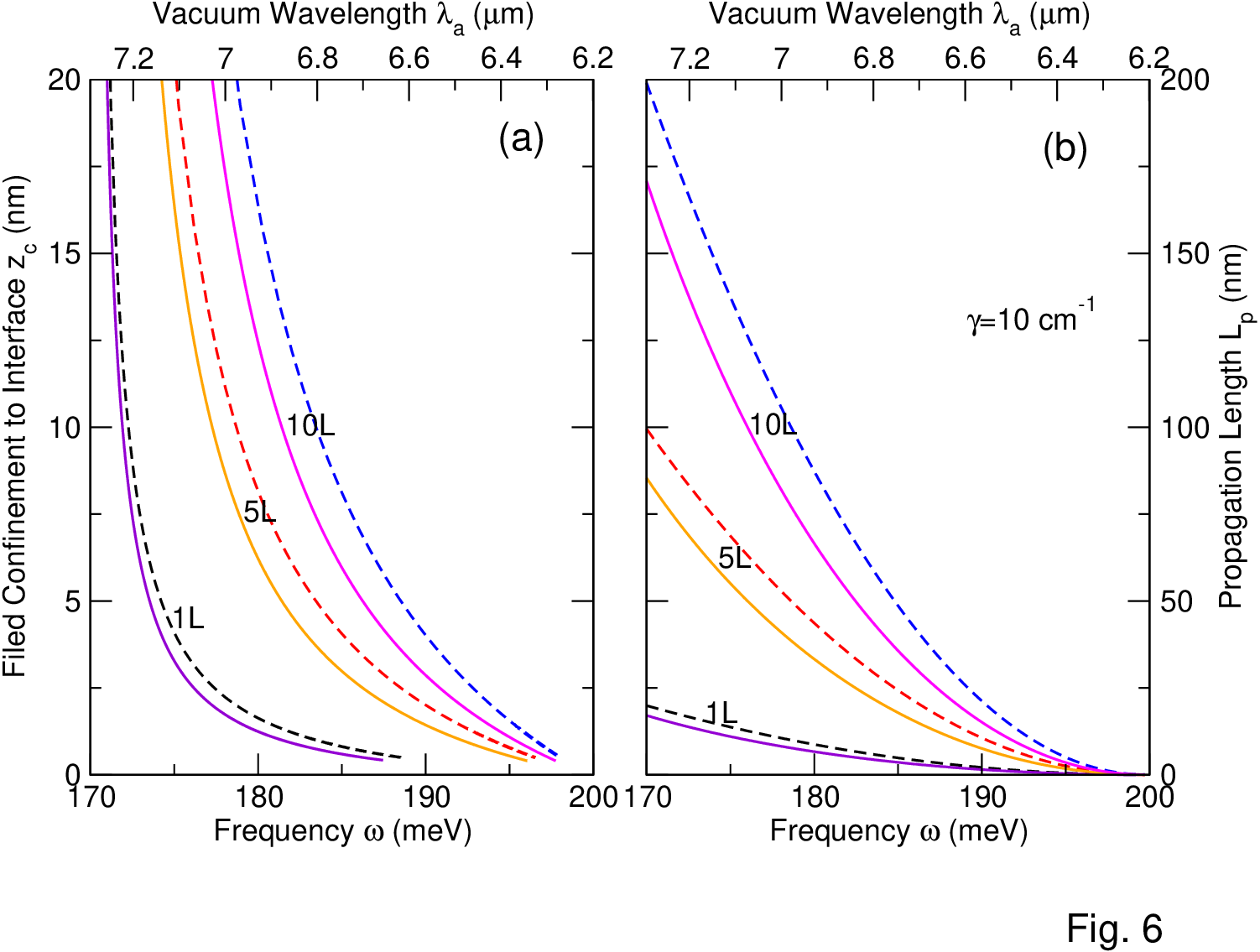}
	\caption
	{(Color online) (a) Field confinement (i.e. decay lengths $z_c$) perpendicular to the interface  and (b) propagation lengths $L_p$ calculated with phonon damping rate $\gamma=$10 cm$^{-1}$ versus PhP frequency $\omega$ (lower horizontal axis) or its corresponding vacuum wavelength $\lambda_a$ (upper horizontal axix), in ML (1L), five-layer (5L) and ten-layer (10L) hBN on SiO$_2$ (solid lines) as well as the corresponding freestanding hBN layers (dashed lines).    
	}
	\label{fig6}
\end{figure}

\begin{figure}
	\includegraphics*[width=12cm]{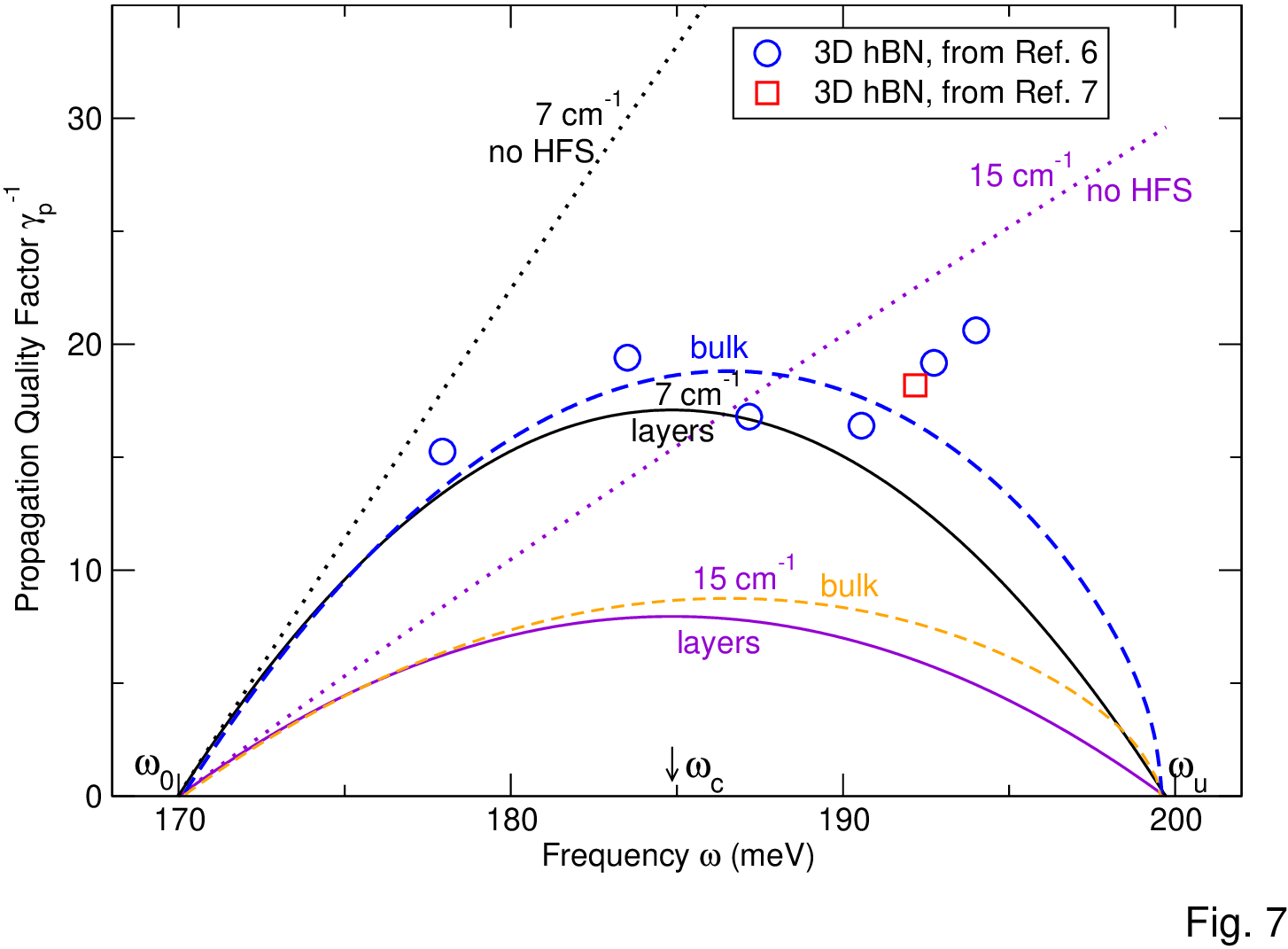}
	\caption
	{(Color online) Propagation quality factors versus PhP frequency for two phonon damping rates $\gamma=$7 and 15 cm$^{-1}$ calculated with (solid lines) or without [dotted lines; see also Fig.~2 (b) of Ref.\cite{Rivera:2019}] accounting for HFS for  $\mathcal{N}$-layer hBN and bulk hBN on SiO$_2$ (dashed lines). Also shown are the experimental data of bulk hBN  on SiO$_2$ (circles from Fig.~5a of Ref.\cite{Giles:2018} for naturally abundant material and square from Ref.\cite{Dai:2014}). 
		The minimum and maximum frequencies $\omega_0$ and $\omega_u$ and centre frequency $\omega_c$ of the PhP frequency band are indicated. 
	}
	\label{fig7}
\end{figure}

\begin{figure}
	\includegraphics*[width=12cm]{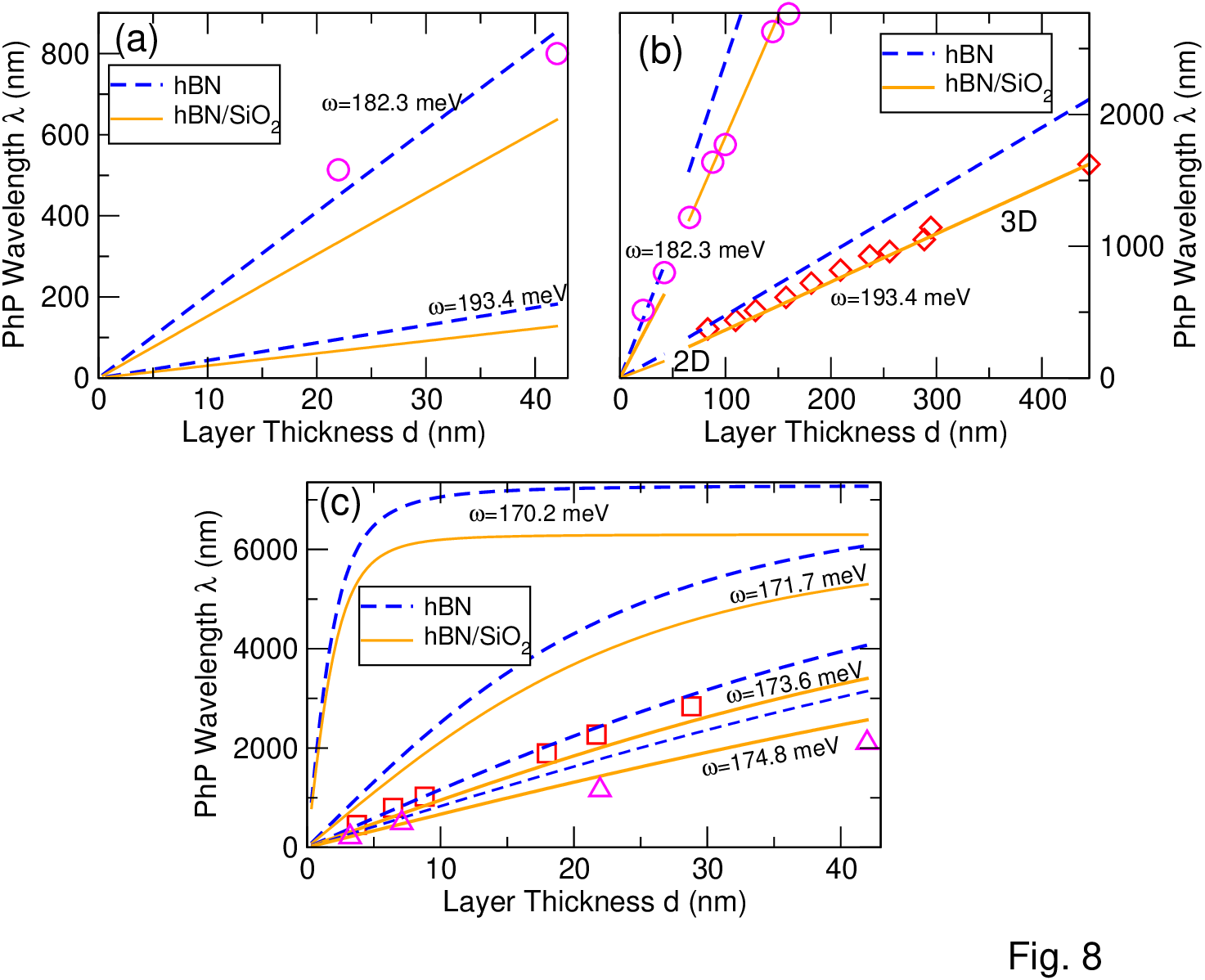}
	\caption
	{(Color online) PhP wavelength $\lambda$ versus layer thickness $d$ in freestanding (dashed lines) and on-SiO$_2$ (solid lines) hBN layers for (a) incident frequencies $\omega$=193.4 and 182.3 meV,  and (c) three  lower frequencies $\omega$=174.8, 173.6 and 171.7 meV and also a frequency near $\omega_0$, $\omega$=170.2 meV.  
		(b) shows the results for $\omega$=193.4 and 182.3 meV calculated from Eq.~(1) of Ref.\cite{Dai:2014} for bulk hBN $d\ge$ 65 nm and with Eq.~(\ref{polaton0}) for hBN layers $d\le$ 42 nm (dashed lines: freestanding hBN; solid lines: on-SiO$_2$ hBN). 
		The symbols are experimental data from s-SNOM measurements for hBN on SiO$_2$:  
		the diamonds and squares are from Fig.~3E ($\omega$=1560 cm$^{-1}$=193.4 meV)  
		and the inset of Fig.~3E ($\omega$=1400 cm$^{-1}$=173.6 meV) of
		Ref.\cite{Dai:2014}, and the circles and triangles are from Fig.~6C ($\omega$=1470 cm$^{-1}$=182.3 meV and $\omega$=1410 cm$^{-1}$=174.8 meV) of
		Ref.\cite{Shiz:2015}, respectively.
	}
	\label{fig8}
\end{figure}

The PhP group velocity, $v_g=d\omega/dk=cdy/dx$, in a freestanding $\mathcal{N}$-layer can be  obtained from Eq.~(\ref{pofre5ap}), 
\begin{equation}
v_g=c\frac{x}{y}\left[1+\frac{\eta}{2(\pi \mathcal{N}\alpha)^2}\frac{y^2-1}{(1+\eta-y^2)^3}\right]^{-1}.     
\label{povg1app}
\end{equation}
 $v_g=c$ as $\omega\rightarrow\omega_0$.  
In the ESA, the $y^2$ term in Eq.~(\ref{pofre5ap}) is dropped and the group velocity reduces to that of the LO modes 
\begin{equation}
	v_g=v_l\mathcal{N}[(1+\eta-y^2)/\eta]^2/y,     
	\label{lovg1app}
\end{equation}
which is proportional to layer number $\mathcal{N}$ for a given frequency (shown in Fig.~\ref{fig3}).   


For $\mathcal{N}$-layer hBN in dielectrics with $\epsilon_1(\omega)$ (substrate) and $\epsilon_2$ (e.g. air), the PhP group velocity is derived after obtaining derivative $dx/dy$ from Eq.~(\ref{pofre2ap}), 
\begin{equation}
v_g=\frac{cx/(2y)[\mathcal{A}(1+\eta-y^2)-\epsilon_2(\bar{\epsilon}_1-\epsilon_2)y^2(y^2-1)/p_2^3]}{p_a+\{\bar{\epsilon}_1^{\prime}p_1-\frac{1}{2}\epsilon_2[\bar{\epsilon}_1^{\prime}\frac{y^2}{p_2}+(\bar{\epsilon}_1-\epsilon_2)\frac{x^2}{p_2^3}]\}(y^2-1)+\mathcal{A}p_b},   
\label{povgdi2app}
\end{equation}
where $\mathcal{A}=4\pi \mathcal{N}\alpha$, $\bar{\epsilon}_1=\bar{\epsilon}_1(y^2)$,  $p_1=\sqrt{x^2-\bar{\epsilon}_1y^2}$, $p_2=\sqrt{x^2-\epsilon_2y^2}$,   $p_a=\bar{\epsilon}_1p_1+\epsilon_2p_1^2/p_2$, and  $p_b=p_1^2+(1+\eta-y^2)(\bar{\epsilon}_1+\bar{\epsilon}_1^{\prime}y^2)/2$.  Like Eq.~(\ref{taylor3ap}), Eq.~(\ref{povgdi2app}) is also independent of the specific form of $\epsilon_1(\omega)$. Given an $\omega$, $x$ in expression~(\ref{povgdi2app}) is found by the numerical solution of Eq.~(\ref{pofre2ap}). The group velocity of Eq.~(\ref{povgdi2app}) at $\omega=\omega_0$ is equal to that of the PhPs of the substrate 
$v_g=c\sqrt{\epsilon_1(\omega_0)}/[\epsilon_1(\omega_0)+\omega_0\epsilon_1^{\prime}(\omega_0)/2]$. The group velocity from the ESA is that of the LO phonons  
\begin{equation}
v_g=\frac{c\mathcal{A}/(2y)}{\frac{\eta(\bar{\epsilon}_1+\epsilon_2)}{(1+\eta-y^2)^2}+\frac{\bar{\epsilon}_1^{\prime}(y^2-1)}{1+\eta-y^2}}.    
\label{lovgdi2app}
\end{equation}

Knowing $v_g$, the PhP density of states is calculated by the relation $g_p(\omega)=k/(2\pi v_g)$. Here the dielectric response $\epsilon_1(\omega)$ is included, as the constant $\epsilon_1$ approximation gives an inaccurate evaluation of $v_g$ and DOS $g_p$ (section III B).  
 As $\omega\rightarrow\omega_0$, the PhP DOS approaches the 2D DOS of photons $g_p(\omega_0)=\omega_0/(2\pi c^2)$ for freestanding layers and the PhP DOS of the substrate, $g_p(\omega_0)=\omega_0/(2\pi c^2)[\epsilon_1(\omega_0)+\omega_0\epsilon_1^{\prime}(\omega_0)/2]$ for  $\mathcal{N}$-layers on SiO$_2$.



\vspace*{10mm}


\vspace*{10mm}

\newpage

\vspace*{20mm}

\vspace*{30cm}

\newpage

\vspace*{10mm}

\newpage

\vspace*{20cm}

\vspace*{30cm}

\newpage

\vspace*{30cm}
\newpage

\vspace*{10mm}

\vspace*{30cm}
\newpage

\vspace*{10mm}

\vspace*{30cm}
\newpage

\vspace*{10mm}

\begin{figure}
	\includegraphics*[width=10cm]{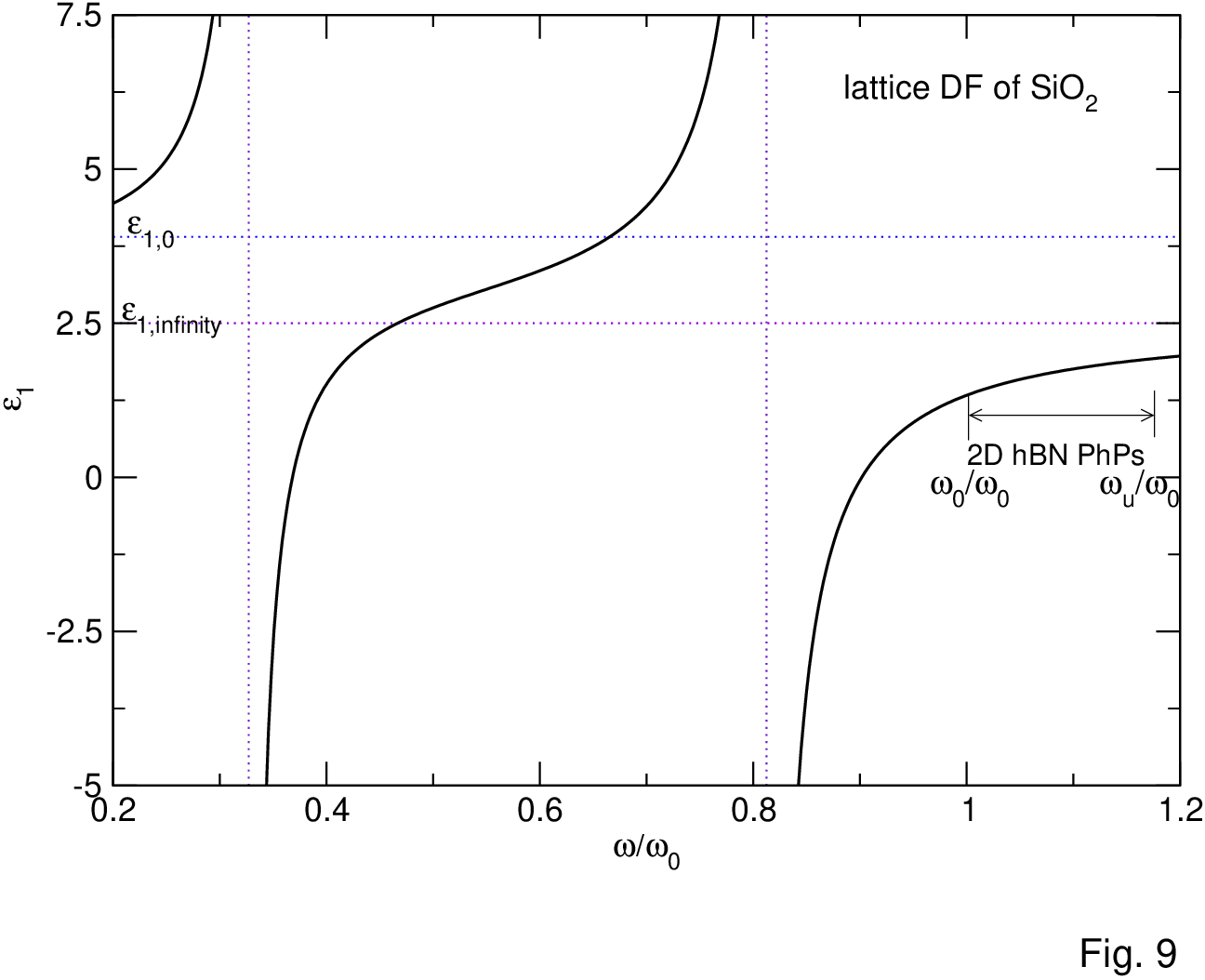}
	\caption
	{(Color online) Lattice dielectric function (DF) of SiO$_2$ $\epsilon_1$ versus $\omega/\omega_0$ [refer to Eq.~(\ref{ldfsio2})], $\omega_0$ being the zone-centre TO and LO phonon frequency of hBN layers; $\epsilon_1$ varies between 1.3 and 1.9 across the 2D hBN PhP frequency band as drawn. The two vertical lines correspond to the two poles $\omega_{T1}$ and $\omega_{T2}$ of the DF and the two horizontal lines indicate the high-frequency and the static DC $\epsilon_{1,\infty}$ and $\epsilon_{1,0}$.  }
	\label{fig9}
\end{figure}

\end{document}